\documentclass[11pt,a4paper]{article}
\usepackage[english]{babel}
\usepackage{amsmath,amsthm,amssymb,epsfig,latexsym}
\usepackage{color}
\usepackage{ulem}
\usepackage[numbers,square,sort&compress]{natbib}

\newcommand{\so}{\scriptscriptstyle \rm I}
\newcommand{\st}{\scriptscriptstyle \rm I\hspace{-1pt}I}
\newcommand{\sth}{\scriptscriptstyle \rm I\hspace{-1pt}I\hspace{-1pt}I}
\newcommand{\la}{u}
\newcommand{\muu}{v}
\newcommand{\lac}{u^{\scriptscriptstyle C}}
\newcommand{\lab}{u^{\scriptscriptstyle B}}
\newcommand{\muc}{v^{\scriptscriptstyle C}}
\newcommand{\mub}{v^{\scriptscriptstyle B}}
\newcommand{\as}{\lambda}
\newcommand{\bla}{\bar u}
\newcommand{\bmu}{\bar v}
\newcommand{\blac}{\bar{u}^{\scriptscriptstyle C}}
\newcommand{\blab}{\bar{u}^{\scriptscriptstyle B}}
\newcommand{\bmuc}{\bar{v}^{\scriptscriptstyle C}}
\newcommand{\bmub}{\bar{v}^{\scriptscriptstyle B}}
\newcommand{\blacp}{\bar{\tilde u}^{\scriptscriptstyle C}}
\newcommand{\blabp}{\bar{\tilde u}^{\scriptscriptstyle B}}
\newcommand{\bmucp}{\bar{\tilde v}^{\scriptscriptstyle C}}
\newcommand{\bmubp}{\bar{\tilde v}^{\scriptscriptstyle B}}
\newcommand{\tz}{\tilde z}
\newcommand{\ta}{\tilde a}
\newcommand{\tb}{\tilde b}

\newcommand{\bet}{\bar\eta}
\newcommand{\bal}{\bar\alpha}
\newcommand{\bw}{\bar w}
\newcommand{\bbet}{\bar\beta}
\newcommand{\bxi}{\bar\xi}
\newcommand{\bg}{\bar\gamma}

\def\Izer{{\sf K}}
\newcommand{\Lot}{\mathcal{L}^{(1,2)}}
\newcommand{\Mot}{\mathcal{M}^{(1,2)}}
\newcommand{\Not}{\mathcal{N}^{(1,2)}}
\newcommand{\Hot}{\mathcal{H}^{(1,2)}}
\newcommand{\Ltt}{\mathcal{L}^{(2,3)}}
\newcommand{\Mtt}{\mathcal{M}^{(2,3)}}
\newcommand{\Ntt}{\mathcal{N}^{(2,3)}}
\newcommand{\Htt}{\mathcal{H}^{(2,3)}}
\newcommand{\tNot}{\widetilde{\mathcal{N}}^{(1,2)}}
\newcommand{\be}[1]{\begin{equation}\label{#1}}
\newcommand{\ba}[1]{\begin{multline}\label{#1}}
\newcommand{\ee}{\end{equation}}
\newcommand{\ea}{\end{eqnarray}}

\newcommand{\num}{\\\rule{0pt}{20pt}}

\newcommand{\tr}{\mathop{\rm tr}}

\def\Izer{{\sf K}}

\newtheorem{prop}{Proposition}[section]
\newtheorem{lemma}{Lemma}[section]
\newtheorem{cor}{Corollary}[section]

\def\qed{\hfill\nobreak\hbox{$\square$}\par\medbreak}
 \makeatletter
 \@addtoreset{equation}{section}
 \makeatother
 
\newcommand{\bea}{\begin{eqnarray}}
\newcommand{\eea}{\end{eqnarray}}

\begin{document}

\begin{flushright}
LAPTH-070/12
\end{flushright}

\vspace{20pt}

\begin{center}
\begin{LARGE}
{\bf Form factors in quantum integrable models with $GL(3)$-invariant $R$-matrix }
\end{LARGE}

\vspace{40pt}

\begin{large}
{S.~Pakuliak${}^a$, E.~Ragoucy${}^b$, N.~A.~Slavnov${}^c$\footnote{pakuliak@theor.jinr.ru, eric.ragoucy@lapth.cnrs.fr, nslavnov@mi.ras.ru}}
\end{large}

 \vspace{12mm}

\vspace{4mm}

${}^a$ {\it Laboratory of Theoretical Physics, JINR, 141980 Dubna, Moscow reg., Russia,\\
Moscow Institute of Physics and Technology, 141700, Dolgoprudny, Moscow reg., Russia,\\
Institute of Theoretical and Experimental Physics, 117259 Moscow, Russia}

\vspace{4mm}

${}^b$ {\it Laboratoire de Physique Th\'eorique LAPTH, CNRS and Universit\'e de Savoie,\\
BP 110, 74941 Annecy-le-Vieux Cedex, France}

\vspace{4mm}

${}^c$ {\it Steklov Mathematical Institute,
Moscow, Russia}

\end{center}


\vspace{4mm}


\begin{abstract}
We study  integrable models solvable by the nested algebraic Bethe ansatz and possessing
$GL(3)$-invariant $R$-matrix.
We obtain determinant representations for form factors of off-diagonal entries of the
monodromy matrix. These representations can be used for the calculation of form factors
and correlation functions of the XXX $SU(3)$-invariant Heisenberg chain.
\end{abstract}

\vspace{1cm}

\vspace{2mm}

\section{Introduction}

The calculation of  correlation functions in quantum integrable models is a very important and complex problem.
A form factor approach is one of the most effective methods for solving this problem.
For this reason, the study of form factors of local operators has attracted the attention of many authors.
There are different methods to address the problem of the calculation of form factors.
In the integrable models of the quantum field theory there exists the `form factor bootstrap approach'
\cite{KarW78,Smi92b,CarM90,Mus92,FriMS90,KouM93,AhnDM93}. It is based on a set of form factors axioms \cite{Smi92b},
which represent a set of difference equations that define specific analytic properties of form factors.
These equations can be solved to provide the integral  representations for form factors. The form factor bootstrap program
is closely related to the approach based on the conformal field theory and its perturbation \cite{Zam91,LukZ97,Luk99,LukZ01}.
A purely  algebraic method to calculate form factors in the infinite chain spin models
was developed by the  Kyoto group \cite{JimMMN92,JimM95L,JimM96}
using the representation theory of quantum affine algebras. This approach also yields
integral formulas for the  form factors of the local operators in such models.
An alternative way to calculate form factors in the spin chain models was developed
by the Lyon group after the  inverse scattering problem was solved and local operators in the spin chain models
were expressed in terms of the monodromy matrix elements \cite{KitMT99}. In this  framework
one can obtain determinant formulas for the form factors of local spin operators. These determinant representations
appeared to be very effective for the calculation of correlation functions \cite{KitKMST11,KitKMST12,CauM05}.

In this article we try to address this problem and continue
our study of form factors in $GL(3)$-invariant quantum integrable models
solvable  by the algebraic Bethe ansatz \cite{FadST79,FadT79,BogIK93L,FadLH96,KulR83}. More precisely,
we calculate matrix elements of the monodromy matrix entries $T_{ij}(z)$ with $|i-j|=1$ between on-shell Bethe vectors
(that is, the eigenstates of the transfer matrix). Recently, in the work \cite{BelPRS13a}, we obtained  determinant representations for form factors of the diagonal elements $T_{ii}(z)$ ($i=1,2,3$) of the monodromy matrix.
Our method was based on the use of a twisted monodromy matrix
\cite{Kor82,IzeK84,KitMST05}. However this approach fails if we deal with form factors of
off-diagonal matrix elements. In this last case, one has to apply a more general method, which
is based on the explicit calculation of the action of the monodromy matrix entries onto  Bethe
vectors. As we have shown in \cite{BelPRS12c} this action gives a linear combination of
 Bethe vectors. Then the  resulting scalar products can be evaluated in terms of sums over
partitions of Bethe parameters \cite{Res86}.

The form factors of the monodromy matrix entries play a very important role.
For a wide class of models, for which
the inverse scattering problem can be solved \cite{KitMT99,MaiT00}, such matrix elements can
be easily associated with form factors of local operators \cite{KitMT99}. In particular, if
$E^{\alpha,\beta}_m$,
$\alpha,\beta=1,2,3$, is an elementary
unit  ($\left(E^{\alpha,\beta}\right)_{jk}=\delta_{j\alpha}\delta_{k\beta}$)
 associated with the $m$-th site of the $SU(3)$-invariant XXX Heisenberg
chain, then
 \be{gen-sol-T}
 E^{\alpha,\beta}_m =(\tr T(0))^{m-1}  T_{\beta\alpha}(0)(\tr T(0))^{-m}.
 \ee
Since the action of the transfer matrix $\tr T(0)$ on on-shell Bethe vectors  is trivial, we see that the form factors of $E^{\alpha,\beta}_m$ are proportional to those of $T_{\beta\alpha}$. Thus, if we have an explicit and compact representations
for form factors of $T_{ij}$, we can study the problem of two-point and multi-point correlation functions, expanding them into series with respect to the form factors.

We have mentioned already that the problem considered in this paper is closely related to the calculation of
Bethe vectors scalar products. In these scalar products, one of the vectors is on-shell, while the other one is off-shell (that is, this vector generically is not an eigenstate of the transfer matrix). A determinant representation
for such type of scalar product was obtained in \cite{Sla89} for $GL(2)$-based models. This representation
allows one to obtain various determinant formulas for form factors. Unfortunately, so far
an analog of this determinant formula is not known in the case of integrable models based on the
$GL(3)$ symmetry. In our previous publications \cite{BelPRS12b,BelPRS13a} we argued that
such an analog hardly exists for the scalar products involving on-shell Bethe vector and a generic off-shell Bethe vector.
However, calculating the form factors of the operators $T_{ij}$ we obtain scalar products involving very specific
off-shell Bethe vectors. For such particular cases of scalar products we succeed to find representations in terms of
a determinant, which is analogous to the determinant representation of \cite{Sla89}.

The article is organized as follows. In section~\ref{S-N} we introduce the model under consideration and describe
the notation used in the paper. We also give there explicit formulas for the scalar product of Bethe vectors
obtained in \cite{Res86} and explain the relationship between different form factors. In section~\ref{S-res} we formulate the main results of this paper.
Section~\ref{S-DER} is devoted to the derivation of the determinant representation for the form factor of the operator $T_{12}$.  In section~\ref{S-OFF} we prove the results for  form factors of other operators $T_{ij}$ with $|i-j|=1$. Appendix~\ref{A-ID} contains the properties of the partition function of the six-vertex model with domain wall boundary conditions and several summation identities for it.

\section{General background\label{S-N}}

\subsection{Generalized $GL(3)$-invariant model}

The models considered below are described by a $GL(3)$-invariant
$R$-matrix acting in the tensor product of two auxiliary spaces $V_1\otimes V_2$, where
$V_k\sim\mathbb{C}^3$, $k=1,2$:
 \be{R-mat}
 R(x,y)=\mathbf{I}+g(x,y)\mathbf{P},\qquad g(x,y)=\frac{c}{x-y}.
 \ee
In the above definition, $\mathbf{I}$ is the identity matrix in $V_1\otimes V_2$, $\mathbf{P}$ is the permutation matrix
that exchanges $V_1$ and $V_2$, and $c$ is a constant.

The monodromy matrix $T(w)$ satisfies the algebra
\be{RTT}
R_{12}(w_1,w_2)T_1(w_1)T_2(w_2)=T_2(w_2)T_1(w_1)R_{12}(w_1,w_2).
\ee
Equation \eqref{RTT} holds in the tensor product $V_1\otimes V_2\otimes\mathcal{H}$,
where $V_k\sim\mathbb{C}^3$, $k=1,2$, are the auxiliary linear spaces, and $\mathcal{H}$ is the Hilbert space of the Hamiltonian of the model under consideration. The  matrices $T_k(w)$ act non-trivially in
$V_k\otimes \mathcal{H}$.

The trace in the auxiliary space $V\sim\mathbb{C}^3$ of the monodromy matrix, $\tr T(w)$, is called the transfer matrix. It is a generating
functional of integrals of motion of the model. The eigenvectors of the transfer matrix are
called on-shell Bethe vectors (or simply on-shell vectors). They can be parameterized by sets of complex parameters
satisfying  Bethe equations (see section~\ref{S-N}).

\subsection{Notations}
We use the same notations and conventions as in the paper \cite{BelPRS13a}.
Besides the function $g(x,y)$ we also introduce a function $f(x,y)$
\be{univ-not}
 f(x,y)=\frac{x-y+c}{x-y}.
\ee
Two other auxiliary functions  will be also used
\be{desand}
h(x,y)=\frac{f(x,y)}{g(x,y)}=\frac{x-y+c}{c},\qquad  t(x,y)=\frac{g(x,y)}{h(x,y)}=\frac{c^2}{(x-y)(x-y+c)}.
\ee
The following obvious properties of the functions introduced above are useful:
 \be{propert}
 g(x,y)=-g(y,x),\quad h(x-c,y)=g^{-1}(x,y),\quad  f(x-c,y)=f^{-1}(y,x),\quad  t(x-c,y)=t(y,x).
 \ee

Before giving a description of the Bethe vectors we formulate a convention on the notations.
We denote sets of variables by bar: $\bar w$, $\bla$, $\bmu$ etc.
Individual elements
of the sets are denoted by subscripts: $w_j$, $\la_k$ etc. Notation $\bar x+c$ means that the
constant $c$ is added to all the elements of the set $\bar x$.
Subsets of variables are denoted by roman indices: $\bla_{\so}$, $\bmu_{\rm iv}$, $\bar w_{\st}$ etc.
In particular, the notation $\bla\Rightarrow\{\bla_{\so},\;\bla_{\st}\}$ means that the
set $\bla$ is divided into two disjoint subsets $\bla_{\so}$ and $\bla_{\st}$.
We assume that the elements in every subset of variables are ordered in such a way that the sequence of
their subscripts is strictly increasing. We call this ordering  natural order.

In order to avoid too cumbersome formulas we use shorthand notations for products of
functions depending on one or two variables. Namely, if functions $g$, $f$, $h$, $t$, as well as $r_1$ and $r_3$ (see \eqref{ratios}) depend on sets of variables, this means that one should take the product over the corresponding set.
For example,
 \be{SH-prod}
 r_1(\bla)=\prod_{\la_j\in\bla} r_1(\la_j);\quad
 g(z, \bar w)= \prod_{w_j\in\bar w} g(z, w_j);\quad
 f(\bla_{\st},\bla_{\so})=\prod_{\la_j\in\bla_{\st}}\prod_{\la_k\in\bla_{\so}} f(\la_j,\la_k).
 \ee
In the last equation of \eqref{SH-prod} the set $\bar u$ is divided into two subsets
$\bla_{\so}$,\; $\bla_{\st}$, and the double product is taken with respect to all
$u_k$ belonging to $\bla_{\so}$ and all $u_j$ belonging to $\bla_{\st}$.
We emphasize once more that this convention is only valid in the case of functions, which are by definition dependent on one or two variables. It does not apply to functions that depend on  sets of variables.

One of the central object in the study of scalar products of $GL(3)$ invariant models is the partition function of the six-vertex model with domain wall boundary conditions (DWPF) \cite{Kor82,Ize87}. We denote it by
$\Izer_n(\bar x|\bar y)$. It depends on two sets of variables $\bar x$ and $\bar y$; the subscript shows that
$\#\bar x=\#\bar y=n$. The function $\Izer_n$ has the following determinant representation \cite{Ize87}
\begin{equation}\label{K-def}
\Izer_n(\bar x|\bar y)
=\Delta'_n(\bar x)\Delta_n(\bar y)h(\bar x,\bar y)
\det_n t(x_j,y_k),
\end{equation}
where $\Delta'_n(\bar x)$ and $\Delta_n(\bar y)$ are
\be{def-Del}
\Delta'_n(\bar x)
=\prod_{j<k}^n g(x_j,x_k),\qquad {\Delta}_n(\bar y)=\prod_{j>k}^n g(y_j,y_k).
\ee
It is easy to see that $\Izer_n$ is symmetric over $\bar x$ and symmetric over $\bar y$, however  $\Izer_n(\bar x|\bar y)\ne
 \Izer_n(\bar y|\bar x)$. Below we consider
$\Izer_n$ depending on combinations of sets of different variables, for example $\Izer_{n}(\bar\xi|\{\bar \alpha,\bar \beta+c\})$.
Due to the symmetry properties of DWPF $\Izer_{n}(\bar\xi|\{\bar \alpha,\bar \beta+c\})=\Izer_{n}(\bar\xi|\{\bar \beta+c,\bar \alpha\})$.

\subsection{Bethe vectors}

Now we pass to the description of Bethe vectors.
A generic Bethe vector is denoted by $\mathbb{B}^{a,b}(\bla;\bmu)$.
It is parameterized by two sets of
complex parameters $\bla=\la_1,\dots,\la_a$ and $\bmu=\muu_1,\dots,\muu_b$ with $a,b=0,1,\dots$.
Dual Bethe vectors are denoted by $\mathbb{C}^{a,b}(\bla;\bmu)$. They also depend on two sets of
complex parameters $\bla=\la_1,\dots,\la_a$ and $\bmu=\muu_1,\dots,\muu_b$. The state with
$\bla=\bmu=\emptyset$ is called a pseudovacuum vector $|0\rangle$. Similarly the dual state
with $\bla=\bmu=\emptyset$ is called a dual pseudovacuum vector $\langle0|$. These vectors
are annihilated by the operators $T_{ij}(w)$, where $i>j$ for  $|0\rangle$ and $i<j$ for $\langle0|$.
At the same time both vectors are eigenvectors for the diagonal entries of the monodromy matrix
 \be{Tjj}
 T_{ii}(w)|0\rangle=\as_i(w)|0\rangle, \qquad   \langle0|T_{ii}(w)=\as_i(w)\langle0|,
 \ee
where $\as_i(w)$ are some scalar functions. In the framework of the generalized model, $\as_i(w)$ remain free functional parameters. Actually, it is always possible to normalize
the monodromy matrix $T(w)\to \as_2^{-1}(w)T(w)$ so as to deal only with the ratios
 \be{ratios}
 r_1(w)=\frac{\as_1(w)}{\as_2(w)}, \qquad  r_3(w)=\frac{\as_3(w)}{\as_2(w)}.
 \ee

If the parameters $\bla$ and $\bmu$ of a Bethe vector\footnote{%
For simplicity here and below we do not distinguish between vectors and dual vectors.}
satisfy a special system of equations (Bethe equations), then
it becomes an eigenvector of the transfer matrix (on-shell Bethe vector). The system of Bethe equations can be written in the following form:
\be{AEigenS-1}
\begin{aligned}
r_1(\bla_{\so})&=\frac{f(\bla_{\so},\bla_{\st})}{f(\bla_{\st},\bla_{\so})}f(\bmu,\bla_{\so}),\\
r_3(\bmu_{\so})&=\frac{f(\bmu_{\st},\bmu_{\so})}{f(\bmu_{\so},\bmu_{\st})}f(\bmu_{\so},\bla).
\end{aligned}
\ee
These equations should hold for arbitrary partitions of the sets $\bla$ and $\bmu$ into subsets
$\{\bla_{\so},\;\bla_{\st}\}$ and $\{\bmu_{\so},\;\bmu_{\st}\}$ respectively. Obviously, it is enough
to demand that the system  \eqref{AEigenS-1} is valid for the
particular case when the sets $\bla_{\so}$ and $\bmu_{\so}$ consist of only one element. Then
\eqref{AEigenS-1} coincides with the standard form of Bethe equations \cite{KulR83}.

If $\bla$ and $\bmu$ satisfy the system \eqref{AEigenS-1}, then
\be{Left-act}
\tr T(w)\mathbb{B}^{a,b}(\bla;\bmu) = \tau(w|\bla,\bmu)\,\mathbb{B}^{a,b}(\bla;\bmu),\qquad
\mathbb{C}^{a,b}(\bla;\bmu)\tr T(w) = \tau(w|\bla,\bmu)\,\mathbb{C}^{a,b}(\bla;\bmu),
\ee
where
\be{tau-def}
\tau(w)\equiv\tau(w|\bla,\bmu)=r_1(w)f(\bla,w)+f(w,\bla)f(\bmu,w)+r_3(w)f(w,\bmu).
\ee

\subsection{Scalar products and form factors}

The scalar products of Bethe vectors are defined as
 \be{SP-def}
 \mathcal{S}_{a,b}\equiv\mathcal{S}_{a,b}(\blac,\bmuc|\blab,\bmub)=
 \mathbb{C}^{a,b}(\blac;\bmuc)\mathbb{B}^{a,b}(\blab;\bmub).
 \ee
We use here superscripts $B$ and $C$ in order to distinguish the sets of parameters
entering these two vectors. In other words, unless explicitly
specified, the variables $\{\blab; \bmub\}$ in $\mathbb{B}^{a,b}$ and
$\{\blac; \bmuc\}$ in $\mathbb{C}^{a,b}$ are not supposed to be  related.

Before giving an explicit formula for the scalar product we introduce the notion of
 highest coefficient $Z_{a,b}(\bar t;\bar x|\bar s; \bar y)$. This function depends
on four sets of variables with cardinalities
$\#\bar t=\#\bar x=a$,  $\#\bar s=\#\bar y=b$, and $a,b=0,1,\dots$. There exist several
explicit representations for the highest coefficient in terms of DWPF \cite{Whe12,BelPRS12a}.
In this paper we use two of them. The first one reads
 \be{RHC-IHC}
  Z_{a,b}(\bar t;\bar x|\bar s; \bar y)=(-1)^b\sum
 \Izer_b(\bar s-c|\bar w_{\so})\Izer_a(\bar w_{\st}|\bar t)
  \Izer_b(\bar y|\bar w_{\so})f(\bar w_{\so},\bar w_{\st}).
    \ee
Here $\bar w=\{\bar s,\;\bar x\}$. The sum is taken with respect to all partitions of the set $\bar w$ into
subsets $\bar w_{\so}$ and $\bar w_{\st}$ with $\#\bar w_{\so}=b$ and $\#\bar w_{\st}=a$.

The second representation has the following form:
 \be{Al-RHC-IHC}
  Z_{a,b}(\bar t;\bar x|\bar s; \bar y)=(-1)^bf(\bar y,\bar x)f(\bar s,\bar t)
   \sum
 \Izer_b(\bal_{\st}-c|\bar y+c)\Izer_a(\bar x|\bal_{\so})\Izer_b(\bal_{\st}-c|\bar s)f(\bal_{\so},\bal_{\st}).
    \ee
Here $\bal=\{\bar y+c,\;\bar t\}$. The sum is taken with respect to all partitions of the set $\bal$ into
subsets $\bal_{\so}$ and $\bal_{\st}$ with $\#\bal_{\so}=a$ and $\#\bal_{\st}=b$.

The scalar product \eqref{SP-def} is a bilinear combination of the highest coefficients. It was calculated in the work \cite{Res86}
 \begin{multline}\label{Resh-SP}
\mathcal{S}_{a,b}(\blac,\bmuc|\blab,\bmub)=\sum r_1(\blab_{\so})r_1(\blac_{\st})r_3(\bmub_{\so}) r_3(\bmuc_{\st})
  f(\blac_{\so},\blac_{\st})  f(\blab_{\st},\blab_{\so})f(\bmuc_{\st},\bmuc_{\so})
   f(\bmub_{\so},\bmub_{\st})
  \num
 \times  \frac{f(\bmuc_{\so},\blac_{\so})f(\bmub_{\st},\blab_{\st})}
 {f(\bmuc,\blac)f(\bmub,\blab)} \;Z_{a-k,n}(\blac_{\st};\blab_{\st}|\bmuc_{\so};\bmub_{\so})
 Z_{k,b-n}(\blab_{\so};\blac_{\so}|\bmub_{\st};\bmuc_{\st}).
 \end{multline}
Here the sum is taken over the partitions of the sets $\blac$, $\blab$, $\bmuc$, and $\bmub$:
 \be{part-1}
 \begin{array}{ll}
 \blac\Rightarrow\{\blac_{\so},\;\blac_{\st}\}, &\qquad  \bmuc\Rightarrow\{\bmuc_{\so},\;\bmuc_{\st}\},\\
 \blab\Rightarrow\{\blab_{\so},\;\blab_{\st}\}, &\qquad  \bmub\Rightarrow\{\bmub_{\so},\;\bmub_{\st}\} .
 \end{array}
 \ee
The partitions are independent except that $\#\blab_{\so}=\#\blac_{\so}=k$ with $k=0,\dots,a$, and $\#\bmub_{\so}=\#\bmuc_{\so}=n$
with $n=0,\dots,b$.

In this formula the parameters $\blac$, $\blab$, $\bmuc$, and $\bmub$ are arbitrary complex numbers, that is $\mathbb{B}^{a,b}(\blab;\bmub)$ and $\mathbb{C}^{a,b}(\blac;\bmuc)$ are generic Bethe vectors. If one of these
vectors, say $\mathbb{C}^{a,b}(\blac;\bmuc)$, is on-shell, then the parameters $\blac$ and $\bmuc$ satisfy
the Bethe equations. In this case one can express the products $r_1(\blac_{\st})$ and $r_3(\bmuc_{\st})$
in terms of the function $f$ via \eqref{AEigenS-1}.

Form factors of the monodromy matrix entries are defined as
 \be{SP-deFF-gen}
 \mathcal{F}_{a,b}^{(i,j)}(z)\equiv\mathcal{F}_{a,b}^{(i,j)}(z|\blac,\bmuc;\blab,\bmub)=
 \mathbb{C}^{a',b'}(\blac;\bmuc)T_{ij}(z)\mathbb{B}^{a,b}(\blab;\bmub),
 \ee
where both $\mathbb{C}^{a',b'}(\blac;\bmuc)$ and $\mathbb{B}^{a,b}(\blab;\bmub)$ are on-shell
Bethe vectors, and
\be{apabpb}
\begin{array}{l}
a'=a+\delta_{i1}-\delta_{j1},\\
b'=b+\delta_{j3}-\delta_{i3}.
\end{array}
\ee
The parameter $z$ is an arbitrary complex  number. Acting with the operator $T_{ij}(z)$ on $\mathbb{B}^{a,b}(\blab;\bmub)$
via formulas obtained in \cite{BelPRS12c} we reduce the form factor to a linear combination of
scalar products, in which $\mathbb{C}^{a',b'}(\blac;\bmuc)$ is on-shell vector.

\subsection{Relations between form factors}

Obviously, there exist nine form factors of $T_{ij}(z)$ in the models with $GL(3)$-invariant
$R$-matrix. However, not all of them are independent. In particular, due to the invariance of the
$R$-matrix under transposition with respect to both spaces, the mapping
\be{def-psi}
\psi\,:
T_{ij}(u) \quad\mapsto\quad T_{ji}(u)
\ee
defines an antimorphism of the algebra \eqref{RTT}. Acting on the Bethe vectors this antimorphism maps them
into the dual ones and vice versa
\be{psiBV}
\psi \big(\mathbb{B}^{a,b}(\bar u;\bar v)\big) = \mathbb{C}^{a,b}(\bar u;\bar v),
\qquad \psi \big(\mathbb{C}^{a,b}(\bar u;\bar v)\big) = \mathbb{B}^{a,b}(\bar u;\bar v).
\ee
Therefore we have
\be{psiFF}
\psi \big(\mathcal{F}_{a,b}^{(i,j)}(z|\blac,\bmuc;\blab,\bmub)\big) =
\mathcal{F}_{a',b'}^{(j,i)}(z|\blab,\bmub;\blac,\bmuc),
\ee
and hence, the
form factor $\mathcal{F}_{a,b}^{(i,j)}(z)$ can be obtained from  $\mathcal{F}_{a,b}^{(j,i)}(z)$
via the replacements $\{\blac,\bmuc\}\leftrightarrow\{\blab,\bmub\}$ and $\{a,b\}\leftrightarrow\{a',b'\}$.

One more relationship
between form factors arises due to the mapping $\varphi$:
\be{def-phi}
\varphi\,:
T_{ij}(u) \quad\mapsto\quad T_{4-j,4-i}(-u),
\ee
that defines an isomorphism of the algebra \eqref{RTT} \cite{BelPRS12c}.
This isomorphism implies the following transform of Bethe vectors:
\be{phiBV}
\varphi \big(\mathbb{B}^{a,b}(\bar u;\bar v)\big) = \mathbb{B}^{b,a}(-\bar v;-\bar u),
\qquad \varphi \big(\mathbb{C}^{a,b}(\bar u;\bar v)\big) = \mathbb{C}^{b,a}(-\bar v;-\bar u).
\ee
Since the mapping $\varphi$ connects the operators $T_ {11}$ and $T_ {33}$, it also leads to the replacement of functions
$r_1 \leftrightarrow r_3$.
Therefore, if $\mathbb{B}^{a,b}(\bar u;\bar v)$ and $\mathbb{C}^{a,b}(\bar u;\bar v)$ are constructed in the representation
$\mathcal{V}\big(r_1(u),r_3(u)\big)$, when their images are in the
representation $\mathcal{V}\big(r_3(-u),r_1(-u)\big)$.  Thus,
\be{phiFF}
\varphi \big(\mathcal{F}_{a,b}^{(i,j)}(z|\blac,\bmuc;\blab,\bmub)\big) =
\mathcal{F}_{b,a}^{(4-j,4-i)}(-z|-\bmuc,-\blac;-\bmub,-\blab)\Bigr|_{r_1\leftrightarrow r_3}.
\ee

\section{Main results\label{S-res}}

The main example considered in this paper is the form factor $ \mathcal{F}_{a,b}^{(1,2)}(z)$:
 \be{SP-deFF}
 \mathcal{F}_{a,b}^{(1,2)}(z)\equiv\mathcal{F}_{a,b}^{(1,2)}(z|\blac,\bmuc;\blab,\bmub)=
 \mathbb{C}^{a',b'}(\blac;\bmuc)T_{12}(z)\mathbb{B}^{a,b}(\blab;\bmub),
 \ee
where both $\mathbb{C}^{a',b'}(\blac;\bmuc)$ and $\mathbb{B}^{a,b}(\blab;\bmub)$ are on-shell
Bethe vectors and $a'=a+1$, $b'=b$.

In order to describe the determinant representation for this form factor we first of all introduce a
set of variables $\bar x=\{x_1,\dots,x_{a'+b}\}$ as the union of three sets
\be{Set-x}
\bar x=\{\blab,z,\bmuc\}=\{\lab_1,\dots,\lab_a,z,\muc_1,\dots,\muc_b\}\,,
\ee
and define a scalar function $\Hot_{a',b}$ as
\be{Hab}
\Hot_{a',b}=\frac{h(\bar x,\blab)h(\bmuc,\bar x)}{h(\bmuc,\blab)}\;\Delta'_{a'}(\blac)
\Delta'_{b}(\bmub)\Delta_{a'+b}(\bar x).
\ee
Then for general $a$ and $b$ we introduce an $a'\times(a'+b)$ matrix
$\Lot(x_k,\lac_j)$ as
\be{Lu-m}
\Lot(x_k,\lac_j)=(-1)^{a'-1}t(\lac_j,x_k)\frac{r_1(x_k)h(\blac,x_k)}{f(\bmuc,x_k)h(x_k,\blab)}
+t(x_k,\lac_j)\frac{h(x_k,\blac)}{h(x_k,\blab)},
\ee
and a $b\times (a'+b)$ matrix
$\Mot(x_k,\mub_j)$ as
\be{Lv-m}
\Mot(x_k,\mub_j)=(-1)^{b-1}t(x_k,\mub_j)\frac{r_3(x_k)h(x_k,\bmub)}{f(x_k,\blab)h(\bmuc,x_k)}
+t(\mub_j,x_k)\frac{h(\bmub,x_k)}{h(\bmuc,x_k)}.
\ee

\begin{prop}\label{PFF-12}
The form factor $\mathcal{F}_{a,b}^{(1,2)}(z)$ admits the following determinant representation:
\be{FF-ans-m}
\mathcal{F}_{a,b}^{(1,2)}(z)=\Hot_{a',b}\;
\det_{a'+b}\Not \,,
\ee
where
\be{Lu-uu-m}
\begin{aligned}
&\Not_ {j,k}=\Lot(x_k,\lac_j),&\qquad j=1,\dots,a',\\
&\Not_ {a'+j,k}=\Mot(x_k,\mub_j),&\qquad  j=1,\dots,b.\\
\end{aligned}
\ee
\end{prop}
The proof of this Proposition is given section~\ref{S-DER}.

{\sl Remark 1}. The order of the elements in the set $\bar x$ is not essential, because the prefactor
$\Delta_{a'+b}(\bar x)$ and $\det_{a'+b}\Not $ are antisymmetric under permutations
of any two elements of $\bar x$. We used the ordering as in \eqref{Set-x}, because it is more convenient for
 the derivation of the determinant representation \eqref{FF-ans-m}

{\sl Remark 2}. It is straightforward to check that due to \eqref{tau-def} the entries of the matrix $\Not $ are proportional to the
Jacobians of the transfer matrix eigenvalues
\be{Lu-mt}
\Lot(x_k,\lac_j)=\frac{c}{f(x_k,\blab)f(\bmuc,x_k)}\frac{g(x_k,\blab)}{g(x_k,\blac)}
\frac{\partial\tau(x_k|\blac,\bmuc)}{\partial\lac_j},
\ee
\be{Lv-mt}
\Mot(x_k,\mub_j)=\frac{-c}{f(x_k,\blab)f(\bmuc,x_k)}\frac{g(\bmuc,x_k)}{g(\bmub,x_k)}
\frac{\partial\tau(x_k|\blab,\bmub)}{\partial\mub_j}.
\ee
In this sense the representation \eqref{FF-ans-m} is an analog of the determinant representations for form factors
in the $GL(2)$-based models \cite{KitMT99}. In particular, at $b=0$ the equation \eqref{FF-ans-m} reproduces the result of
\cite{KitMT99}.

Determinant representations for other form factors $ \mathcal{F}_{a,b}^{(i,j)}(z)$ with $|i-j|=1$ can be
derived from \eqref{FF-ans-m} by  the mappings \eqref{psiFF}, \eqref{phiFF}. First, we give  the explicit
formulas for the form factor of the operator $T_{23}$
 \be{3SP-deFF}
 \mathcal{F}_{a,b}^{(2,3)}(z)\equiv\mathcal{F}_{a,b}^{(2,3)}(z|\blac,\bmuc;\blab,\bmub)=
 \mathbb{C}^{a',b'}(\blac;\bmuc)T_{23}(z)\mathbb{B}^{a,b}(\blab;\bmub),
 \ee
where $a'=a$ and $b'=b+1$.

We introduce  a set of variables $\bar y=\{y_1,\dots,y_{a+b'}\}$ as the union of three sets
\be{3Set-x}
\bar y=\{\bmub,z,\blac\}=\{\mub_1,\dots,\mub_b,z,\lac_1,\dots,\lac_a\}\,,
\ee
and a function
\be{3Hab}
\Htt_{a,b'}=\frac{h(\bar y,\blac)h(\bmub,\bar y)}{h(\bmub,\blac)}\;\Delta'_{a}(\blab)
\Delta'_{b'}(\bmuc)\Delta_{a+b'}(\bar y).
\ee
Then for general $a$ and $b$ we define an $a\times(a+b')$ matrix
$\Ltt(y_k,\lab_j)$ as
\be{3Lu-m}
\Ltt(y_k,\lab_j)=(-1)^{a-1}t(\lab_j,y_k)\frac{r_1(y_k)h(\blab,y_k)}{f(\bmub,y_k)h(y_k,\blac)}
+t(y_k,\lab_j)\frac{h(y_k,\blab)}{h(y_k,\blac)},
\ee
and a $b'\times (a+b')$ matrix
$\Mtt(y_k,\muc_j)$ as
\be{3Lv-m}
\Mtt(y_k,\muc_j)=(-1)^{b'-1}t(y_k,\muc_j)\frac{r_3(y_k)h(y_k,\bmuc)}{f(y_k,\blac)h(\bmub,y_k)}
+t(\muc_j,y_k)\frac{h(\bmuc,y_k)}{h(\bmub,y_k)}.
\ee

\begin{prop}\label{PFF-23}
The form factor $\mathcal{F}_{a,b}^{(2,3)}(z)$ admits the following determinant representation:
\be{3FF-ans-m}
\mathcal{F}_{a,b}^{(2,3)}(z)=\Htt_{a,b'}\;
\det_{a+b'}\Ntt\,,
\ee
where
\be{3Lu-uu-m}
\begin{aligned}
&\Ntt_{j,k}=\Ltt(y_k,\lab_j),&\qquad j=1,\dots,a,\\
&\Ntt_{a+j,k}=\Mtt(y_k,\muc_j),&\qquad  j=1,\dots,b'.\\
\end{aligned}
\ee
\end{prop}


Similarly to the case considered in Proposition~\ref{PFF-12}, the order of the elements in the set $\bar y$ is not essential,
and the entries of the matrix $\Ntt $ can be expressed in terms of the
Jacobians of the transfer matrix eigenvalues.

\begin{prop}\label{PFF-32-21}
The form factor $\mathcal{F}_{a,b}^{(3,2)}(z)$ admits the following determinant representation:
\be{32FF-ans-m}
\mathcal{F}_{a,b}^{(3,2)}(z)=\Hot_{a',b}\;
\det_{a'+b}\Not\,,
\ee
where $\Hot_{a',b}$ and $\Not$ are given by \eqref{Hab} and \eqref{FF-ans-m} respectively.

The form factor $\mathcal{F}_{a,b}^{(2,1)}(z)$ admits the following determinant representation:
\be{21FF-ans-m}
\mathcal{F}_{a,b}^{(2,1)}(z)=\Htt_{a,b'}\;
\det_{a+b'}\Ntt\,,
\ee
where $\Htt_{a,b'}$ and $\Ntt$ are given by \eqref{3Hab} and \eqref{3FF-ans-m} respectively.
\end{prop}

The proofs of  Proposition~\ref{PFF-23} and Proposition~\ref{PFF-32-21} are given section~\ref{S-OFF}.

{\sl Remark}. We would like to stress that although the representations \eqref{32FF-ans-m} and \eqref{21FF-ans-m}
formally coincide with \eqref{FF-ans-m} and \eqref{3FF-ans-m}, the values of $a'$ and $b'$ in these formulas are different.
Indeed, one has $a'=a+1$ and $b'=b$ in \eqref{FF-ans-m}, while $a'=a$ and $b'=b-1$ in \eqref{32FF-ans-m}. Similarly
$a'=a$ and $b'=b+1$ in \eqref{3FF-ans-m}, while $a'=a-1$ and $b'=b$ in \eqref{21FF-ans-m}. Therefore, in particular,
the matrices $\Not$ and $\Ntt$ in \eqref{FF-ans-m} and \eqref{3FF-ans-m} have a size
$(a+b+1)\times(a+b+1)$, while in the equations \eqref{32FF-ans-m} and \eqref{21FF-ans-m} the same
matrices have a size $(a+b)\times(a+b)$.

\section{Derivation of the determinant representation\label{S-DER}}

In this section we prove the determinant representation \eqref{FF-ans-m} for the form factor of the
operator $T_{12}(z)$. We use the same technique as in the work \cite{BelPRS12b}.

First of all we need a formula for the action of $T_{12}$ on the Bethe vectors \cite{BelPRS12c}
 \be{act12}
 T_{12}(z)\mathbb{B}^{a,b}(\bla;\bmu)=-\sum
 f(\bar\xi_{\rm ii},\bar\xi_{\rm i})\Izer_1(\bar\xi_{\rm i}|z+c)\,
 \mathbb{B}^{a+1,b}(\bar\eta;\bar\xi_{\rm ii}).
 \ee
Here $\{\bmu,z\}=\bar\xi$ and $\{\bla,z\}=\bar\eta$.
The sum is taken over partitions  $\bar\xi\Rightarrow\{\bar\xi_{\rm i},\bar\xi_{\rm ii}\}$ with $\#\bar\xi_{\rm i}=1$.
Multiplying  \eqref{act12} from the left by $\mathbb{C}^{a+1,b}(\blac;\bmuc)$ we reduce the form factor $\mathcal{F}_{a,b}^{(1,2)}(z)$ to a linear combination of scalar products
\be{FF12-SP}
\mathcal{F}_{a,b}^{(1,2)}(z)=-\sum
 f(\bar\xi_{\rm ii},\bar\xi_{\rm i})\Izer_1(\bar\xi_{\rm i}|z+c) \;
 \mathcal{S}_{a+1,b}(\blac,\bmuc|\bar\eta,\bar\xi_{\rm ii}).
\ee
Now we can substitute here the expression \eqref{Resh-SP} for the scalar product, replacing there
the set $\blab$ by $\bar\eta$ and the set $\bmub$ by $\bar\xi_{\rm ii}$.
Using the Bethe equations for the set $\blac$
\be{Be-1}
r_1(\blac_{\st})f(\blac_{\so},\blac_{\st})=f(\blac_{\st},\blac_{\so})f(\bmuc,\blac_{\st}),
\ee
we obtain
 \begin{multline}\label{FF-begin}
 \mathcal{F}_{a,b}^{(1,2)}(z)=-\sum f(\bar\xi_{\st},\bar\xi_{\rm i})f(\bar\xi_{\so},\bar\xi_{\rm i})\Izer_1(\bar\xi_{\rm i}|z+c)
  r_1(\bet_{\so})r_3(\bxi_{\so}) r_3(\bmuc_{\st})
    \num
  \times \frac{f(\blac_{\st},\blac_{\so})  f(\bet_{\st},\bet_{\so})f(\bmuc_{\st},\bmuc_{\so})
   f(\bxi_{\so},\bxi_{\st})}
 {f(\bmuc_{\st},\blac_{\so})f(\bxi_{\st},\bet_{\so})f(\bxi_{\so},\bet)} \;Z_{a+1-k,n}(\blac_{\st};\bet_{\st}|\bmuc_{\so};\bxi_{\so})
 Z_{k,b-n}(\bet_{\so};\blac_{\so}|\bxi_{\st};\bmuc_{\st}).
 \end{multline}
The sum is taken with respect to partitions:
 \be{part-2}
 \begin{aligned}
 &\blac\Rightarrow\{\blac_{\so},\;\blac_{\st}\}, &\qquad  &\bmuc\Rightarrow\{\bmuc_{\so},\;\bmuc_{\st}\},\\
 &\bet\Rightarrow\{\bet_{\so},\;\bet_{\st}\}, &\qquad  &\bxi\Rightarrow\{\bar\xi_{\rm i},\;\bxi_{\so},\;\bxi_{\st}\} ,
 \end{aligned}
 \ee
where $\#\bet_{\so}=\#\blac_{\so}=k$ with $k=0,\dots,a+1$; $\#\bar\xi_{\rm i}=1$; and $\#\bxi_{\so}=\#\bmuc_{\so}=n$
with $n=0,\dots,b$.

Substituting here \eqref{RHC-IHC} for $Z_{a+1-k,n}$ and \eqref{Al-RHC-IHC} for $Z_{k,b-n}$ we find
 \begin{multline}\label{FF-Z-K}
 \mathcal{F}_{a,b}^{(1,2)}(z)=\sum (-1)^{b+1}f(\bar\xi_{\so},\bar\xi_{\rm i})
  \frac{r_1(\bet_{\so})r_3(\bxi_{\so}) r_3(\bmuc_{\st})} {f(\bxi_{\so},\bet)}
    f(\bet_{\st},\bet_{\so})f(\bmuc_{\st},\bmuc_{\so})
   f(\bxi_{\so},\bxi_{\st})f(\bw_{\so},\bw_{\st})f(\bal_{\so},\bal_{\st})
    \num
  \times
  \Bigl[\Izer_1(\bar\xi_{\rm i}|z+c)\Izer_{b-n}(\bal_{\st}-c|\bxi_{\st})f(\bar\xi_{\st},\bar\xi_{\rm i})\Bigr]\cdot
  \Bigl[  \Izer_{k}(\blac_{\so}|\bal_{\so})\Izer_{a-k+1}(\bw_{\st}|\blac_{\st})f(\blac_{\st},\blac_{\so})\Bigr]\num
 \times  \;\Izer_{n}(\bmuc_{\so}-c|\bw_{\so}) \Izer_{n}(\bxi_{\so}|\bw_{\so})\Izer_{b-n}(\bal_{\st}-c|\bmuc_{\st}+c).
 \end{multline}
Here $\bw=\{\bet_{\st},\bmuc_{\so}\}$ and $\bal=\{\bet_{\so},\bmuc_{\st}+c\}$. The sum is taken with respect to the partitions
\eqref{part-2} and two additional partitions:
$\bw\Rightarrow\{\bw_{\so},\bw_{\st}\}$ and $\bal\Rightarrow\{\bal_{\so},\bal_{\st}\}$ with $\#\bw_{\so}=n$ and
$\#\bal_{\so}=k$.

{\sl Remark.} Note that the restrictions on the cardinalities of subsets are explicitly specified by the
subscripts of DWPF. For example, the DWPF $\Izer_{k}(\blac_{\so}|\bal_{\so})$
is defined only if $\#\blac_{\so}=\#\bal_{\so}=k$. Therefore below we do not specify the cardinalities
of subsets in separate comments.

Now we can apply \eqref{Sym-Part-old1} to the terms in the square brackets in the second line of \eqref{FF-Z-K}.
The sum with respect to the partitions $\blac\Rightarrow\{\blac_{\so},\blac_{\st}\}$ gives
\be{ML-1}
\sum\Izer_{k}(\blac_{\so}|\bal_{\so})\Izer_{a-k+1}(\bw_{\st}|\blac_{\st})f(\blac_{\st},\blac_{\so})=
(-1)^k f(\blac,\bal_{\so})\Izer_{a+1}(\{\bal_{\so}-c,\bw_{\st}\}|\blac).
\ee
Similarly, setting $\{\bar\xi_{\rm i},\bxi_{\st}\}=\bxi_{\sth}$ we calculate the sum
with respect to the partitions $\bxi_{\sth}\Rightarrow\{\bar\xi_{\rm i},\bxi_{\st}\}$:
\be{ML-2}
\sum  \Izer_1(\bar\xi_{\rm i}|z+c)\Izer_{b-n}(\bal_{\st}-c|\bxi_{\st})f(\bar\xi_{\st},\bar\xi_{\rm i})=
-f^{-1}(z,\bar\xi_{\sth})\Izer_{b-n+1}(\{z,\bal_{\st}-c\}|\bxi_{\sth}),
\ee
where we have used \eqref{propert}.
Then \eqref{FF-Z-K} turns into
 \begin{multline}\label{FF-1st-sum}
 \mathcal{F}_{a,b}^{(1,2)}(z)=\sum (-1)^{b+k}
  \frac{r_1(\bet_{\so})r_3(\bxi_{\so}) r_3(\bmuc_{\st})} {f(z,\bxi_{\sth})f(\bxi_{\so},\bet)}
 f(\bet_{\st},\bet_{\so})f(\bmuc_{\st},\bmuc_{\so}) f(\bxi_{\so},\bxi_{\sth}) f(\bw_{\so},\bw_{\st})
   f(\bal_{\so},\bal_{\st}) \num
  \times f(\blac,\bal_{\so})\;\Izer_{a+1}(\{\bal_{\so}-c,\bw_{\st}\}|\blac)\;\Izer_{n}(\bmuc_{\so}-c|\bw_{\so})\num
 \times\Izer_{b-n+1}(\{z,\bal_{\st}-c\}|\bxi_{\sth}) \Izer_{n}(\bxi_{\so}|\bw_{\so})
\Izer_{b-n}(\bal_{\st}-c|\bmuc_{\st}+c)\, .
 \end{multline}

Now one should distinguish between two cases: $z\in\bxi_{\sth}$ or $z\in\bxi_{\so}$. In the first case the contribution to the form factor
does not depend on $r_3(z)$, while in the second case it is proportional to $r_3(z)$. Thus, we can write
\be{F-OO}
\mathcal{F}_{a,b}^{(1,2)}(z)=r_3(z)\Omega_2+\Omega_1.
\ee
We will calculate $\Omega_1$ and $\Omega_2$ separately.

\subsection{The first particular case}

Here we consider the case $z\in\bxi_{\sth}$. The corresponding contribution $\Omega_1$ to the form factor does
not depend on $r_3(z)$. Therefore without loss of generality below we will set $r_3(z)=0$.

We can set $\bxi_{\so}=\bmub_{\so}$ and $\bxi_{\sth}=\{z,\bmub_{\st}\}$. Then the product $f^{-1}(z,\bxi_{\sth})$ vanishes,
however this zero is compensated by the pole of $\Izer_{b-n+1}(\{z,\bal_{\st}-c\}|\bxi_{\sth})$ (see \eqref{Rec-Ky}):
\be{zero-pole}
f^{-1}(z,\bxi_{\sth})\Izer_{b-n+1}(\{z,\bal_{\st}-c\}|\bxi_{\sth})=
f^{-1}(z,\bal_{\st})\Izer_{b-n}(\bal_{\st}-c|\bmub_{\st}).
\ee
Substituting this into \eqref{FF-1st-sum} and using Bethe equations for $r_3(\bxi_{\so})=r_3(\bmub_{\so})$
we obtain after simple algebra
 \begin{multline}\label{Nor3z-1}
\Omega_1=\sum (-1)^{b+k}
  \frac{r_1(\bet_{\so})r_3(\bmuc_{\st})} {f(z,\bal_{\st})}
f(\blac,\bal_{\so}) f(\bet_{\st},\bet_{\so})
    f(\bmuc_{\st},\bmuc_{\so})f(\bw_{\so},\bw_{\st}) f(\bal_{\so},\bal_{\st}) \;\Izer_{n}(\bmuc_{\so}-c|\bw_{\so})\num
  \times \Izer_{b-n}(\bal_{\st}-c|\bmuc_{\st}+c) \Izer_{a+1}(\{\bal_{\so}-c,\bw_{\st}\}|\blac)
 \cdot\Bigl[\Izer_{n}(\bmub_{\so}|\bw_{\so})  \Izer_{b-n}(\bal_{\st}-c|\bmub_{\st}) f(\bmub_{\st},\bmub_{\so})\Bigr].
  \end{multline}
The sum with respect to the partitions $\bmub\Rightarrow\{\bmub_{\so},\bmub_{\st}\}$ (see the terms in the square brackets
in \eqref{Nor3z-1}) can be calculated via \eqref{Sym-Part-old1}
\be{ML-3}
\sum\Izer_{n}(\bmub_{\so}|\bw_{\so})  \Izer_{b-n}(\bal_{\st}-c|\bmub_{\st}) f(\bmub_{\st},\bmub_{\so})=
(-1)^n f(\bmub,\bw_{\so})\;\Izer_{b}(\{\bw_{\so}-c,\bal_{\st}-c\}|\bmub).
\ee
Thus, we obtain
 \begin{multline}\label{Nor3z-1st-sum}
\Omega_1=\sum (-1)^{b+k+n}
  \frac{r_1(\bet_{\so})r_3(\bmuc_{\st})} {f(z,\bal_{\st})}
f(\blac,\bal_{\so})f(\bmub,\bw_{\so}) f(\bet_{\st},\bet_{\so})
    f(\bmuc_{\st},\bmuc_{\so})f(\bw_{\so},\bw_{\st}) f(\bal_{\so},\bal_{\st})\num
  \times
 \;\Izer_{n}(\bmuc_{\so}-c|\bw_{\so})
 \Izer_{b-n}(\bal_{\st}-c|\bmuc_{\st}+c) \Izer_{b}(\{\bal_{\st}-c,\bw_{\so}-c\}|\bmub)
 \Izer_{a+1}(\{\bal_{\so}-c,\bw_{\st}\}|\blac).
 \end{multline}

Now it is necessary to specify the partitions of the sets $\bal$ and $\bw$. We set
\be{New-part}
\begin{aligned}
&\bal_{\so}=\{\bet_{\rm ii},\bmuc_{\rm i}+c\},\qquad &\bet_{\so}=\{\bet_{\rm i},\bet_{\rm ii}\},\\
&\bal_{\st}=\{\bet_{\rm i},\bmuc_{\rm ii}+c\},\qquad &\bet_{\st}=\{\bet_{\rm iii},\bet_{\rm iv}\},\\
&{}&{}\\
&\bw_{\so}=\{\bet_{\rm iv},\bmuc_{\rm iii}\},\qquad &\bmuc_{\so}=\{\bmuc_{\rm iii},\bmuc_{\rm iv}\},\\
&\bw_{\st}=\{\bet_{\rm iii},\bmuc_{\rm iv}\},\qquad &\bmuc_{\st}=\{\bmuc_{\rm i},\bmuc_{\rm ii}\}.
\end{aligned}
\ee
We denote the cardinalities of these subsets as $\#\bet_j=k_j$ and $\#\bmuc_j=n_j$, where $j={\rm i, ii, iii, iv}$. Evidently
$\sum_{j={\rm i}}^{\rm iv}k_j=a+1$ and $\sum_{j={\rm i}}^{\rm iv}n_j=b$. It is also easy to see that
\be{card-new-part}
\begin{aligned}
&n_{\rm i}+n_{\rm ii}=b-n,\qquad &k_{\rm i}+k_{\rm ii}=k,\\
&n_{\rm iii}+n_{\rm iv}=n,\qquad &k_{\rm iii}+k_{\rm iv}=a+1-k,\\
&n_{\rm i}=k_{\rm i},\qquad &n_{\rm iv}=k_{\rm iv}.
\end{aligned}
\ee
In terms of the subsets introduced above the equation \eqref{Nor3z-1st-sum} takes the form
 \begin{multline}\label{sum-New-ss}
\Omega_1=\sum (-1)^{b+k+n}
  \frac{r_1(\bet_{\rm i})r_1(\bet_{\rm ii})r_3(\bmuc_{\rm i})r_3(\bmuc_{\rm ii})} {f(z,\bet_{\rm i})f(\bmuc_{\rm i},\blac)}
 f(\bmuc_{\rm ii},z)f(\blac,\bet_{\rm ii})f(\bmub,\bet_{\rm iv}) f(\bmub,\bmuc_{\rm iii})\num
 \times f(\bet_{\rm iii},\bet_{\rm i})f(\bet_{\rm iii},\bet_{\rm ii})f(\bet_{\rm iv},\bet_{\rm i})f(\bet_{\rm iv},\bet_{\rm ii})
f(\bet_{\rm iv},\bet_{\rm iii})f(\bet_{\rm ii},\bet_{\rm i})\;f(\bmuc_{\rm i},\bmuc_{\rm iii})f(\bmuc_{\rm i},\bmuc_{\rm iv})\num
\times f(\bmuc_{\rm ii},\bmuc_{\rm iii})f(\bmuc_{\rm ii},\bmuc_{\rm iv})
f(\bmuc_{\rm i},\bmuc_{\rm ii})f(\bmuc_{\rm iii},\bmuc_{\rm iv})\;
 f(\bet_{\rm iv},\bmuc_{\rm iv})f(\bmuc_{\rm iii},\bet_{\rm iii})\frac{f(\bmuc_{\rm i}+c,\bet_{\rm i})}{f(\bmuc_{\rm ii},\bet_{\rm ii})}\num
  \times
   \Izer_{n}(\{\bmuc_{\rm iii}-c,\bmuc_{\rm iv}-c\}|\{\bet_{\rm iv},\bmuc_{\rm iii}\})\;
   \Izer_{b-n}(\{\bet_{\rm i}-c,\bmuc_{\rm ii}\}|\{\bmuc_{\rm i}+c,\bmuc_{\rm ii}+c\})\num
  \times
 \Izer_{b}(\{\bet_{\rm i}-c,\bet_{\rm iv}-c,\bmuc_{\rm iii}-c,\bmuc_{\rm ii}\}|\bmub)\;
 \Izer_{a+1}(\{\bet_{\rm ii}-c,\bet_{\rm iii},\bmuc_{\rm i},\bmuc_{\rm iv}\}|\blac).
 \end{multline}

Now several simplifications are possible. First of all the DWPF $\Izer_n$ and $\Izer_{b-n}$ can be transformed
via \eqref{K-K}, \eqref{Red-K}
\be{IHC-tr-1}
\Izer_{n}(\{\bmuc_{\rm iii}-c,\bmuc_{\rm iv}-c\}|\{\bet_{\rm iv},\bmuc_{\rm iii}\})=
(-1)^n\Izer_{n_{\rm iv}}(\bet_{\rm iv}|\bmuc_{\rm iv})f^{-1}(\bet_{\rm iv},\bmuc_{\rm iv}),
\ee
\be{IHC-tr-2}
\Izer_{b-n}(\{\bet_{\rm i}-c,\bmuc_{\rm ii}\}|\{\bmuc_{\rm i}+c,\bmuc_{\rm ii}+c\}) =(-1)^{b-n}
\Izer_{n_{\rm i}}(\bmuc_{\rm i}+c|\bet_{\rm i})f^{-1}(\bmuc_{\rm i}+c,\bet_{\rm i}).
\ee

Then one should express $r_1(\bet_{\rm i})$ and $r_3(\bmuc_{\rm i})$ in terms of the Bethe equations. Observe
that $z\notin\bet_{\rm i}$, due to the factor $f^{-1}(z,\bet_{\rm i})$. Therefore the subset $\bet_{\rm i}$ consists
of the elements $\lab_j$ only, and one do can use the Bethe equations for $r_1(\bet_{\rm i})$. Therefore
\be{r1-BE}
r_1(\bet_{\rm i})=\frac{f(\bet_{\rm i},\bet_{\rm ii})f(\bet_{\rm i},\bet_{\rm iii})f(\bet_{\rm i},\bet_{\rm iv})f(z,\bet_{\rm i})}
{f(\bet_{\rm ii},\bet_{\rm i})f(\bet_{\rm iii},\bet_{\rm i})f(\bet_{\rm iv},\bet_{\rm i})f(\bet_{\rm i},z)}f(\bmub,\bet_{\rm i}),
\ee
and
\be{r3-BE}
r_3(\bmuc_{\rm i})=\frac{f(\bmuc_{\rm ii},\bmuc_{\rm i})f(\bmuc_{\rm iii},\bmuc_{\rm i})f(\bmuc_{\rm iv},\bmuc_{\rm i})}
{f(\bmuc_{\rm i},\bmuc_{\rm ii})f(\bmuc_{\rm i},\bmuc_{\rm iii})f(\bmuc_{\rm i},\bmuc_{\rm iv})}
f(\bmuc_{\rm i},\blac).
\ee
These expressions should be substituted into \eqref{sum-New-ss}.

{\sl Remark.} Formally one can also use the Bethe equations for the product $r_3(\bmuc_{\rm ii})$.
However it is more convenient to keep this product as it is.

Finally, we introduce new subsets
\be{NNew-part}
\begin{aligned}
&\bet_{\so}=\{\bet_{\rm i},\bet_{\rm iv}\},\qquad &\bmuc_{\so}=\{\bmuc_{\rm i},\bmuc_{\rm iv}\},\\
&\bet_{\st}=\{\bet_{\rm ii},\bet_{\rm iii}\},\qquad &\bmuc_{\st}=\{\bmuc_{\rm ii},\bmuc_{\rm iii}\},
\end{aligned}
\ee
and we denote $n_{\so}=\#\bet_{\so}=\#\bmuc_{\so}$. We draw the readers attention that these new subsets
have nothing to do with the subsets used in \eqref{Nor3z-1st-sum}. We use, however, the
same notation, as we deal with the sum over partitions, and therefore it does not matter how we denote separate terms
of this sum.

Then the equation \eqref{sum-New-ss} can be written in the
following form:
 \begin{equation}\label{sum-NNew-ss}
\Omega_1=\sum (-1)^{b}
 f(\bmuc_{\st},\bet_{\st})f(\bet_{\so},\bet_{\st})f(\bmuc_{\st},\bmuc_{\so}){\sf G}_{n_{\so}}(\bet_{\so}|\bmuc_{\so})
 {\sf L}_{a+1}(\{\bet_{\st},\bmuc_{\so}\}|\blac) {\sf M}_{b}(\{\bet_{\so},\bmuc_{\st}\}|\bmub),
 \end{equation}
where we have introduced three new functions: ${\sf G}_{n_{\so}}$, $ {\sf L}_{a+1}$, and  ${\sf M}_{b}$.
Originally all of them are defined as sums over partitions.
The function ${\sf G}_{n_{\so}}$ is given by
\be{G-def}
 {\sf G}_{n_{\so}}(\bet_{\so}|\bmuc_{\so})=\sum \frac{f(\bet_{\rm i},\bet_{\rm iv})f(\bmuc_{\rm iv},\bmuc_{\rm i})}{f(\bet_{\rm i},z)}
 \Izer_{n_{\rm iv}}(\bet_{\rm iv}|\bmuc_{\rm iv})\Izer_{n_{\rm i}}(\bmuc_{\rm i}+c|\bet_{\rm i}),
 \ee
where the sum is taken over partitions $\bet_{\so}\Rightarrow\{\bet_{\rm i},\bet_{\rm iv}\}$
and $\bmuc_{\so}\Rightarrow\{\bmuc_{\rm i},\bmuc_{\rm iv}\}$.

The function ${\sf L}_{a+1}(\{\bet_{\st},\bmuc_{\so}\}|\blac)$ reads
\be{L-def}
 {\sf L}_{a+1}(\{\bet_{\st},\bmuc_{\so}\}|\blac)=\sum\frac{(-1)^{k_{\rm ii}}r_1(\bet_{\rm ii})}{f(\bmuc,\bet_{\rm ii})}
  f(\blac,\bet_{\rm ii})f(\bet_{\rm iii},\bet_{\rm ii})f(\bmuc_{\so},\bet_{\rm ii})
   \Izer_{a+1}(\{\bet_{\rm ii}-c,\bet_{\rm iii},\bmuc_{\so}\}|\blac),
   \ee
where the sum is taken over partitions $\bet_{\st}\Rightarrow\{\bet_{\rm ii},\bet_{\rm iii}\}$.

Finally, the function ${\sf M}_{b}(\{\bet_{\so},\bmuc_{\st}\}|\bmub)$ is given by
\begin{multline}\label{M-def}
  {\sf M}_{b}(\{\bet_{\so},\bmuc_{\st}\}|\bmub)=\sum\frac{(-1)^{n_{\rm ii}}r_3(\bmuc_{\rm ii})}{f(\bmuc_{\rm ii},\blab)}
f(\bmub,\bet_{\so})f(\bmub,\bmuc_{\rm iii})\\
\times f(\bmuc_{\rm ii},\bet_{\so})f(\bmuc_{\rm ii},\bmuc_{\rm iii})
 \Izer_{b}(\{\bet_{\so}-c,\bmuc_{\rm iii}-c,\bmuc_{\rm ii}\}|\bmub),
   \end{multline}
where the sum is taken over partitions
$\bmuc_{\st}\Rightarrow\{\bmuc_{\rm ii},\bmuc_{\rm iii}\}$.
It is straightforward to check that substituting the definitions \eqref{G-def}--\eqref{M-def}
into \eqref{sum-NNew-ss} we reproduce \eqref{sum-New-ss}.

It is remarkable that all the sums with respect to partitions in \eqref{G-def}--\eqref{M-def}
can be explicitly computed.
The function $ {\sf G}_{n_{\so}}(\bet_{\so}|\bmuc_{\so})$ can be calculated via \eqref{Ident-G1}
\be{G-res1}
 {\sf G}_{n_{\so}}(\bet_{\so}|\bmuc_{\so})=(-1)^{n_{\so}}t(\bmuc_{\so},\bet_{\so}) h(\bmuc_{\so},\bmuc_{\so})
 h(\bet_{\so},\bet_{\so})
\frac{h(\bmuc_{\so},z)}{h(\bet_{\so},z)}.
 \ee

The functions ${\sf L}_{a+1}(\{\bet_{\st},\bmuc_{\so}\}|\blac)$ and ${\sf M}_{b}(\{\bet_{\so},\bmuc_{\st}\}|\bmub)$
can be calculated via lemma~\ref{Long-Det}. Let us set in \eqref{SumDet1}
\be{Lem-1}
\bg=\{\bet_{\st},\bmuc_{\so}\}, \qquad C_1(\gamma)=-\frac{r_1(\gamma)}{f(\bmuc,\gamma)},\qquad C_2(\gamma)=1.
\ee
Observe that $C_1(\muc_k)=0$ due to the factor $f^{-1}(\bmuc,\gamma)$. Therefore, dividing in \eqref{SumDet1} the set
$\bg$ into two subsets $\{\bg_{\so},\bg_{\st}\}$ one should consider only the partitions for
which $\bmuc_{\so}\subset\bg_{\st}$. It means that actually we deal with the partitions of
the subset $\bet_{\st}$ only. Namely, we can set $\bg_{\so}=\bet_{\rm ii}$ and  $\bg_{\st}=\{\bmuc_{\so},\bet_{\rm iii}\}$.
Then the sum in \eqref{SumDet1} coincides with the sum \eqref{L-def} and we obtain
\be{L-res}
 {\sf L}_{a+1}(\bg|\blac)=\Delta'_{a+1}(\blac)\Delta_{a+1}(\bg)
\det_{a+1}\Bigr[\Lot(\gamma_k,\lac_j)h(\gamma_k,\blab)\Bigl], \qquad
\bg=\{\bet_{\st},\bmuc_{\so}\},
\ee
where the matrix $\Lot$ is given by \eqref{Lu-m}.

Similarly for the calculation \eqref{M-def} one should set in the sum \eqref{SumDet1}
\be{Lem-2}
\bg=\{\bet_{\so},\bmuc_{\st}\},\qquad C_1(\gamma)=1,\qquad C_2(\gamma)=-\frac{r_3(\gamma)}{f(\gamma,\blab)}.
\ee
Then $C_2(\eta_k)=0$ either due to the product $f^{-1}(\gamma,\blab)$ or due to the condition
$r_3(z)=0$  (that we freely imposed in this subsection). Therefore we can set $\bg_{\so}=\{\bmuc_{\rm iii},\bet_{\so}\}$ and $\bg_{\st}=\bmuc_{\rm ii}$
in \eqref{SumDet1}. Then the sum \eqref{SumDet1} turns into  \eqref{M-def} and we obtain
\be{M-res}
{\sf M}_{b}(\bg|\bmub)=(-1)^b\Delta'_b(\bmub)\Delta_b(\bg)
\det_{b}\Bigr[\Mot(\gamma_k,\mub_j)h(\bmuc,\gamma_k)\Bigl], \qquad
\bg=\{\bet_{\so},\bmuc_{\st}\},
\ee
where the matrix $\Mot$ is given by \eqref{Lv-m}.

Introducing
\be{hLhM}
\hat{\sf L}_{a+1}(\bg|\blac)=\frac{{\sf L}_{a+1}(\bg|\blac)}{h(\bg,\blab)},\qquad
\hat{\sf M}_{b}(\bg|\bmub)=(-1)^b\frac{{\sf M}_{b}(\bg|\bmub)}{h(\bmuc,\bg)},
\ee
we obtain after simple algebra
 \begin{equation}\label{last-sum2}
\Omega_1=\frac{f(\bmuc,\bet)h(\bmuc,\bmuc)h(\bet,\bet)}{h(\bet,z)}
 \sum (-1)^{n_{\so}} \frac{g(\bet_{\so},\bet_{\st})g(\bmuc_{\st},\bmuc_{\so})}
 {g(\bmuc_{\so},\bet_{\st})g(\bmuc_{\st},\bet_{\so})}
\; \hat{\sf L}_{a+1}(\{\bet_{\st},\bmuc_{\so}\}|\blac)
 \hat{\sf M}_{b}(\{\bet_{\so},\bmuc_{\st}\}|\bmub).
 \end{equation}

Define a set $\bar x$ as in \eqref{Set-x}
\be{x-def}
\bar x=\{\bet,\bmuc\}=\{\lab_1,\dots\lab_a,z,\muc_1,\dots,\muc_b\}.
\ee
For arbitrary partition $\bar x\Rightarrow\{\bar x_{\so},\bar x_{\st}\}$ with $\#\bar x_{\so}=a+1$ and
$\#\bar x_{\st}=b$ we have
\begin{equation}\label{id-triv}
1=\frac{\Delta_{a+b+1}(\bar x)}{\Delta_{a+b+1}(\bar x)}
=(-1)^{P_{\so,\st}}\frac{\Delta_{a+1}(\bar x_{\so})\Delta_b(\bar x_{\st})g(\bar x_{\st},\bar x_{\so})}
{\Delta_{a+1}(\bet)\Delta_b(\bmuc)g(\bmuc,\bet)},
\end{equation}
where $P_{\so,\st}$ is the parity of the permutation mapping the sequence $\{\bar x_{\so},\bar x_{\st}\}$ into the
ordered sequence $x_1,\dots,x_{a+b+1}$. Setting
$\bar x_{\so}=\{\bet_{\st},\bmuc_{\so}\}$ and $\bar x_{\st}=\{\bet_{\so},\bmuc_{\st}\}$ we obtain after
elementary algebra
\begin{equation}\label{id-triv1}
1=(-1)^{P_{\so,\st}+n_{\so}}\frac{\Delta_{a+1}(\bar x_{\so})\Delta_b(\bar x_{\st})}
{\Delta_{a+1}(\bet)\Delta_b(\bmuc)} \frac{g(\bet_{\so},\bet_{\st})g(\bmuc_{\st},\bmuc_{\so})}
 {g(\bmuc_{\so},\bet_{\st})g(\bmuc_{\st},\bet_{\so})}.
\end{equation}
Thus, the equation \eqref{last-sum2} can be written in the form
 \begin{multline}\label{last-sum3}
\Omega_1=f(\bmuc,\bet)h(\bmuc,\bmuc)h(\blab,\blab)h(z,\blab)\Delta'_{a+1}(\blac)
 \Delta'_b(\bmub)\Delta_{a+1}(\bet)\Delta_b(\bmuc)\\
\times\sum(-1)^{P_{\so,\st}} \det_{a+1}[\Lot(x^{\so}_k,\lac_j)]\;
\det_{b}[\Mot(x^{\st}_k,\mub_j)].
 \end{multline}
Here $x^{\so}_k$ (resp. $x^{\st}_k$) is the $k$-th element of the subset $x_{\so}$ (resp. $x_{\st}$). It is easy to see that the
prefactor in the first line of \eqref{last-sum3} coincides with the function $\Hot_{a+1,b}$ (see \eqref{Hab}).
The sum \eqref{last-sum3} is nothing but the expansion of the determinant of the $(a+b+1)\times(a+b+1)$ matrix $\Not$
with respect to the first $(a+1)$ rows. Thus, we finally obtain
\be{FF1-ans-m1}
\Omega_1=\Hot_{a+1,b}\;
\det_{a+b+1}\Not \Bigr|_{r_3(z)=0}.
\ee

\subsection{The second particular case}

Now we turn back to the equation \eqref{FF-1st-sum} and consider the case $z\in\bxi_{\so}$, that is we
 compute the term $\Omega_2$ in \eqref{F-OO}. The general idea of the calculation is the same as in
 the case of $\Omega_1$, however there are several subtleties.

Since $\bet=\{z,\blab\}$, the product $f^{-1}(\bxi_{\so},\bet)$ vanishes. The only possible way to obtain
non-zero contribution to $\Omega_2$ is to compensate this zero by the pole of
$\Izer_{n}(\bxi_{\so}|\bw_{\so})$. The last one occurs if and only if $z\in\bw_{\so}$, which implies
$z\in\bet_{\st}$. Thus, we can set
\be{z-in}
\begin{aligned}
&\bxi_{\so}=\{z,\bmub_{\so}\},\qquad &\bxi_{\sth}=\bmub_{\st},\\
&\bw_{\so}=\{z,\bw_{0}\},\qquad &\bw_{\st}=\bw_{\st},\\
&\bet_{\so}=\blab_{\so},\qquad &\bet_{\st}=\{z,\blab_{\st}\}.
\end{aligned}
\ee
Substituting this into \eqref{FF-1st-sum} we obtain
 \begin{multline}\label{r3z-1}
\Omega_2=\sum (-1)^{b+k}
  r_1(\blab_{\so})r_3(\bmuc_{\st})
f(\blac,\bal_{\so}) f(z,\bmuc_{\so}) f(\blab_{\st},\blab_{\so})
    f(\bmuc_{\st},\bmuc_{\so})f(\bw_{0},\bw_{\st}) f(\bal_{\so},\bal_{\st})\num
  \times
\Izer_{n}(\bmuc_{\so}-c|\{\bw_{0},z\})\Izer_{b-n}(\bal_{\st}-c|\bmuc_{\st}+c) \Izer_{a+1}(\{\bal_{\so}-c,\bw_{\st}\}|\blac)\num
 \times
  \Izer_{b-n+1}(\{z,\bal_{\st}-c\}|\bmub_{\st}) \Izer_{n-1}(\bmub_{\so}|\bw_{0}) f(\bmub_{\st},\bmub_{\so}),
 \end{multline}
where we have also used the Bethe equations for $r_3(\bmub_{\so})$:
 \be{BE-mub1}
r_3(\bmub_{\so})=\frac{f(\bmub_{\st},\bmub_{\so})}{f(\bmub_{\so},\bmub_{\st})}f(\bmub_{\so},\blab).
\ee
Applying \eqref{Sym-Part-old1} to the terms in the last line of \eqref{r3z-1} we take the sum over partitions $\bmub\Rightarrow\{\bmub_{\so},\bmub_{\st}\}$:
\be{ML-4}
\sum\Izer_{b-n+1}(z,\bal_{\st}-c|\bmub_{\st}) \Izer_{n-1}(\bmub_{\so}|\bw_{0}) f(\bmub_{\st},\bmub_{\so})=
(-1)^{n-1}f(\bmub,\bw_0)\Izer_{b}(\{\bw_0-c,\bal_{\st}-c,z\}|\bmub).
\ee
Thus, we arrive at
 \begin{multline}\label{r3z-1-sum}
\Omega_2=\sum (-1)^{b+k+n+1}
  r_1(\blab_{\so})r_3(\bmuc_{\st})
f(\blac,\bal_{\so})f(\bmub,\bw_{0})  f(z,\bmuc_{\so}) f(\blab_{\st},\blab_{\so})
    f(\bmuc_{\st},\bmuc_{\so})f(\bw_{0},\bw_{\st}) \num
  \times f(\bal_{\so},\bal_{\st})
 \;\Izer_{n}(\bmuc_{\so}-c|\{\bw_{0},z\})
\Izer_{b-n}(\bal_{\st}-c|\bmuc_{\st}+c)
 %
\Izer_{b}(\{\bw_{0}-c,\bal_{\st}-c,z\}|\bmub)\Izer_{a+1}(\{\bal_{\so}-c,\bw_{\st}\}|\blac).
 \end{multline}

Now we should specify the subsets similarly to \eqref{New-part}
\be{New-part2}
\begin{aligned}
&\bal_{\so}=\{\blab_{\rm ii},\bmuc_{\rm i}+c\},\qquad &\blab_{\so}=\{\blab_{\rm i},\blab_{\rm ii}\},\\
&\bal_{\st}=\{\blab_{\rm i},\bmuc_{\rm ii}+c\},\qquad &\blab_{\st}=\{\blab_{\rm iii},\blab_{\rm iv}\},\\
&{}&{}\\
&\bw_{0}=\{\blab_{\rm iv},\bmuc_{\rm iii}\},\qquad &\bmuc_{\so}=\{\bmuc_{\rm iii},\bmuc_{\rm iv}\},\\
&\bw_{\st}=\{\blab_{\rm iii},\bmuc_{\rm iv}\},\qquad &\bmuc_{\st}=\{\bmuc_{\rm i},\bmuc_{\rm ii}\}.
\end{aligned}
\ee
We again denote the cardinalities of the subsets above as $\#\blab_j=k_j$ and $\#\bmuc_j=n_j$. Now
$\sum_{j={\rm i}}^{\rm iv}k_j=a$, $\sum_{j={\rm i}}^{\rm iv}n_j=b$ and
\be{card-new-part2}
\begin{aligned}
&n_{\rm i}+n_{\rm ii}=b-n,\qquad &k_{\rm i}+k_{\rm ii}=k,\\
&n_{\rm iii}+n_{\rm iv}=n,\qquad &k_{\rm iii}+k_{\rm iv}=a-k,\\
&n_{\rm i}=k_{\rm i},\qquad &n_{\rm iv}=k_{\rm iv}+1.
\end{aligned}
\ee

Using the new subsets we obtain an analog of \eqref{sum-New-ss}
 \begin{multline}\label{sum-New-ss2}
\Omega_2=\sum (-1)^{b+k+n}
  \frac{r_1(\blab_{\rm i})r_1(\blab_{\rm ii})r_3(\bmuc_{\rm i})r_3(\bmuc_{\rm ii})} {f(\bmuc_{\rm i},\blac)}
 f(z,\bmuc_{\rm iii})f(z,\bmuc_{\rm iv})
 f(\blac,\blab_{\rm ii})f(\bmub,\blab_{\rm iv}) f(\bmub,\bmuc_{\rm iii})\num
 \times f(\blab_{\rm iii},\blab_{\rm i})f(\blab_{\rm iii},\blab_{\rm ii})f(\blab_{\rm iv},\blab_{\rm i})f(\blab_{\rm iv},\blab_{\rm ii})
f(\blab_{\rm iv},\blab_{\rm iii})f(\blab_{\rm ii},\blab_{\rm i})\;f(\bmuc_{\rm i},\bmuc_{\rm iii})f(\bmuc_{\rm i},\bmuc_{\rm iv})\num
\times f(\bmuc_{\rm ii},\bmuc_{\rm iii})f(\bmuc_{\rm ii},\bmuc_{\rm iv})
f(\bmuc_{\rm i},\bmuc_{\rm ii})f(\bmuc_{\rm iii},\bmuc_{\rm iv})\;
 f(\blab_{\rm iv},\bmuc_{\rm iv})f(\bmuc_{\rm iii},\blab_{\rm iii})\frac{f(\bmuc_{\rm i}+c,\blab_{\rm i})}{f(\bmuc_{\rm ii},\blab_{\rm ii})}\num
  \times
   \Izer_{n}(\{\bmuc_{\rm iii}-c,\bmuc_{\rm iv}-c\}|\{z,\blab_{\rm iv},\bmuc_{\rm iii}\})\;
   \Izer_{b-n}(\{\blab_{\rm i}-c,\bmuc_{\rm ii}\}|\{\bmuc_{\rm i}+c,\bmuc_{\rm ii}+c\})\num
  \times
 \Izer_{b}(\{\blab_{\rm i}-c,\blab_{\rm iv}-c,\bmuc_{\rm iii}-c,\bmuc_{\rm ii},z\}|\bmub)\;
 \Izer_{a+1}(\{\blab_{\rm ii}-c,\blab_{\rm iii},\bmuc_{\rm i},\bmuc_{\rm iv}\}|\blac).
 \end{multline}

Now one should make the same transforms as before. Namely, we should simplify $\Izer_{n}$
and $\Izer_{b-n}$ via \eqref{K-K}, \eqref{Red-K}; express $r_1(\blab_{\rm i})$ and $r_3(\bmuc_{\rm i})$ in terms of Bethe equations;
introduce new subsets
\be{NNew-part2}
\begin{aligned}
&\blab_{\so}=\{\blab_{\rm i},\blab_{\rm iv}\},\qquad &\bmuc_{\so}=\{\bmuc_{\rm i},\bmuc_{\rm iv}\},\\
&\blab_{\st}=\{\blab_{\rm ii},\blab_{\rm iii}\},\qquad &\bmuc_{\st}=\{\bmuc_{\rm ii},\bmuc_{\rm iii}\}.
\end{aligned}
\ee
Pay attention that now $n_{\so}=\#\blab_{\so}+1=\#\bmuc_{\so}$. We also introduce
 $z'=z+c$. Then the equation \eqref{sum-New-ss2} can be written in the
following form:
 \begin{multline}\label{sum-NNew-ss2}
\Omega_2=f^{-1}(\bmub,z')\sum (-1)^{b+1}
 \frac{f(\bmuc_{\st},\blab_{\st})}{f(\bmuc_{\st},z')}f(\blab_{\so},\blab_{\st})f(\bmuc_{\st},\bmuc_{\so})\\
 \times \widetilde {\sf G}_{n_{\so}}(\blab_{\so}|\bmuc_{\so})
 \widetilde {\sf L}_{a+1}(\{\blab_{\st},\bmuc_{\so}\}|\blac) \widetilde{\sf M}_{b}(\{\blab_{\so},\bmuc_{\st},
 z'\}|\bmub).
 \end{multline}
Here
\be{G-def2}
 \widetilde {\sf G}_{n_{\so}}(\blab_{\so}|\bmuc_{\so})=\sum f(\blab_{\rm i},\blab_{\rm iv})f(\bmuc_{\rm iv},\bmuc_{\rm i})
 \Izer_{n_{\rm iv}}(\{z,\blab_{\rm iv}\}|\bmuc_{\rm iv})\Izer_{n_{\rm i}}(\bmuc_{\rm i}+c|\blab_{\rm i}),
 \ee
where the sum is taken over partitions $\blab_{\so}\Rightarrow\{\blab_{\rm i},\blab_{\rm iv}\}$
and $\bmuc_{\so}\Rightarrow\{\bmuc_{\rm i},\bmuc_{\rm iv}\}$.

The function $ \widetilde{\sf L}_{a+1}(\{\blab_{\st},\bmuc_{\so}\}|\blac)$ is
\be{L-def2}
 \widetilde{\sf L}_{a+1}(\{\blab_{\st},\bmuc_{\so}\}|\blac)=\sum\frac{(-1)^{k_{\rm ii}}r_1(\blab_{\rm ii})}{f(\bmuc,\blab_{\rm ii})}
  f(\blac,\blab_{\rm ii})f(\blab_{\rm iii},\blab_{\rm ii})f(\bmuc_{\so},\blab_{\rm ii})
   \Izer_{a+1}(\{\blab_{\rm ii}-c,\blab_{\rm iii},\bmuc_{\so}\}|\blac),
   \ee
where the sum is taken over partitions $\blab_{\st}\Rightarrow\{\blab_{\rm ii},\blab_{\rm iii}\}$.

The function $\widetilde{\sf M}_{b}(\{\blab_{\so},\bmuc_{\st}, z'\}|\bmub)$ is
\begin{multline}\label{M-def2}
  \widetilde{\sf M}_{b}(\{\blab_{\so},\bmuc_{\st}, z'\}|\bmub)=\sum\frac{(-1)^{n_{\rm ii}}r_3(\bmuc_{\rm ii})}{f(\bmuc_{\rm ii},\blab)}
f(\bmub,\blab_{\so})f(\bmub,\bmuc_{\rm iii})f(\bmub,z')\\
\times f(\bmuc_{\rm ii},\blab_{\so})f(\bmuc_{\rm ii},\bmuc_{\rm iii})f(\bmuc_{\rm ii},z')
 \Izer_{b}(\{\blab_{\so}-c,\bmuc_{\rm iii}-c, z'-c,\bmuc_{\rm ii}\}|\bmub),
   \end{multline}
where the sum is taken over partitions
 $\bmuc_{\st}\Rightarrow\{\bmuc_{\rm ii},\bmuc_{\rm iii}\}$.

The function $\widetilde {\sf G}_{n_{\so}}(\blab_{\so}|\bmuc_{\so})$ can be calculated via
\eqref{Ident-G2}
\be{G-res2}
 \widetilde {\sf G}_{n_{\so}}(\blab_{\so}|\bmuc_{\so})=(-1)^{n_{\so}}
 \frac{t(\bmuc_{\so},\blab_{\so}) h(\bmuc_{\so},\bmuc_{\so})
 h(\blab_{\so},\blab_{\so})}{h(\bmuc_{\so},z')g(z',\blab_{\so})}.
 \ee
The calculation of $ \widetilde{\sf L}_{a+1}(\{\blab_{\st},\bmuc_{\so}\}|\blac)$ is the same as
the one of ${\sf L}_{a+1}(\{\bet_{\st},\bmuc_{\so}\}|\blac)$ (one should only replace everywhere
$\bet_{\st}$ by $\blab_{\st}$).

The calculation of $\widetilde{\sf M}_{b}(\{\blab_{\so},\bmuc_{\st}, z'\}|\bmub)$ also is almost
the same as before. The difference is that now it depends on additional parameter $z'$. However
this difference does not make sense, if we set by definition $r_3(z')=0$. We can always do it,
because the form factor anyway does not depend on $r_3(z')$. Thus, we find
\be{L-res2}
\widetilde{\sf L}_{a+1}(\bg|\blac)=\Delta'_{a+1}(\blac)\Delta_{a+1}(\bg)
\det_{a+1}\Bigr[\Lot(\gamma_k,\lac_j)h(\gamma_k,\blab)\Bigl], \qquad
\bg=\{\blab_{\st},\bmuc_{\so}\},
\ee
and
\be{M-res2}
\widetilde{\sf M}_{b}(\bg|\bmub)=(-1)^b\Delta'_b(\bmub)\Delta_b(\bg)
\det_{b}\Bigr[\Mot(\gamma_k,\mub_j)h(\bmuc,\gamma_k)\Bigl], \qquad
\bg=\{\blab_{\so},z',\bmuc_{\st}\}.
\ee
Formally, the obtained representations coincide with \eqref{L-res}, \eqref{M-res}. However  the sets $\bg$ are
different.
In \eqref{L-res2} the set $\bg$ does not contain $z$, while in \eqref{L-res} it could contain $z$.
Respectively, in \eqref{M-res2} the set $\bg$  contains $z'$, while in \eqref{M-res}
it was $z$-independent.

Introducing $\hat{\sf L}_{a+1}$ and $\hat{\sf M}_{b}$ by
\be{hLhM-t}
\hat{\sf L}_{a+1}(\bg|\blac)=\frac{\widetilde{\sf L}_{a+1}(\bg|\blac)}{h(\bg,\blab)},\qquad
\hat{\sf M}_{b}(\bg|\bmub)=(-1)^b\frac{\widetilde{\sf M}_{b}(\bg|\bmub)}{h(\bmuc,\bg)},
\ee
and substituting
\eqref{G-res2}--\eqref{M-res2} into \eqref{sum-NNew-ss2} we after simple algebra arrive at the
analog of \eqref{last-sum2}
 \begin{multline}\label{2-last-sum2}
\Omega_2=\frac{f(\bmuc,\blab)h(\bmuc,\bmuc)h(\blab,\blab)}{f(\bmub,z')}
 \sum (-1)^{n_{\so}+1} \frac{g(\blab_{\so},\blab_{\st})g(\bmuc_{\st},\bmuc_{\so})}
 {g(\bmuc_{\so},\blab_{\st})g(\bmuc_{\st},\blab_{\so})}\\
\times \frac{\hat{\sf L}_{a+1}(\{\blab_{\st},\bmuc_{\so}\}|\blac)
 \hat{\sf M}_{b}(\{\blab_{\so},\bmuc_{\st},z'\}|\bmub)}{g(\bmuc_{\st},z')g(z',\blab_{\so})}.
 \end{multline}

Similarly to \eqref{x-def} we introduce a set $\bar x'$ as
\be{x-def2}
\bar x'=\{\blab,z',\bmuc\}=\{\lab_1,\dots\lab_a,z',\muc_1,\dots,\muc_b\}.
\ee
Consider partitions $\bar x'\Rightarrow\{\bar x'_{\so},\;\bar x'_{\st}\}$ with $\#\bar x'_{\so}=a$ and
$\#\bar x'_{\st}=b+1$.
One can set $\bar x'_{\so}=\{\blab_{\st},\bmuc_{\so}\}$ and $\bar x'_{\st}=\{\blab_{\so},z',\bmuc_{\st}\}$.
Then the analog of \eqref{id-triv1} has the following form:
\begin{equation}\label{2-id-triv1}
1=(-1)^{P_{\so,\st}}\frac{\Delta_{a+1}(\bar x'_{\so})\Delta_b(\bar x'_{\st})}
{\Delta_a(\blab)\Delta_b(\bmuc)} \frac{g(\blab_{\so},\blab_{\st})g(\bmuc_{\st},\bmuc_{\so})}
 {g(\bmuc_{\so},\blab_{\st})g(\bmuc_{\st},\blab_{\so})}
 \frac{(-1)^{n_{\so}}}{g(\bmuc_{\st},z')g(z',\blab_{\so})}.
\end{equation}
It is important that  unlike the previous case we have $\#\bmuc_{\so}=n_{\so}$
and $\#\blab_{\so}=n_{\so}-1$. Therefore, in particular,
\be{g-g}
g(\blab_{\so},\bmuc_{\so})=g(\bmuc_{\so},\blab_{\so}).
\ee

Thus, the equation \eqref{2-last-sum2} can be written in the form
 \begin{multline}\label{2-last-sum3}
\Omega_2=-\frac{f(\bmuc,\blab)h(\bmuc,\bmuc)h(\blab,\blab)}{f(\bmub,z')}\Delta'_{a+1}(\blac)
 \Delta'_b(\bmub)\Delta_a(\blab)\Delta_b(\bmuc)\\
\times\sum(-1)^{P_{\so,\st}} \det_{a+1}[\Lot({x'}^{\so}_k,\lac_j)]\;
\det_{b}[\Mot({x'}^{\st}_k,\mub_j)],
 \end{multline}
provided $r_3(z')=0$. This formula is almost the expansion of the $(a+b+1)\times(a+b+1)$ determinant
with respect to the first $(a+1)$ rows. We should take care only about the condition $z'\in\bar x'_{\st}$.
This can be done if we set by definition $\Lot(z',\lac_j)\equiv 0$. We do can impose this
constraint, since $\Lot(z',\lac_j)$ does not enter the formula \eqref{2-last-sum3}.
Then we obtain
 \begin{equation}\label{2-last-sum4}
\Omega_2=\frac{-\Hot_{a+1,b}}{f(\bmub,z')f(z,\blab)f(\bmuc,z)}
\;\det_{a'+b}\tNot,\qquad \bar x'=\{\lab_1,\dots\lab_a,z',\muc_1,\dots,\muc_b\},
 \end{equation}
where
\be{Lu-uu-mt}
\begin{aligned}
&\tNot_{j,k}=\Lot(x'_k,\lac_j),&\qquad j=1,\dots,a+1,\quad k\ne a+1,\\
&\tNot_{a+1+j,k}=\Mot(x'_k,\mub_j),&\qquad  j=1,\dots,b,\quad k\ne a+1,\\
&\tNot_{j,a+1}=0,&\qquad j=1,\dots,a+1,\\
&\tNot_{a+1+j,a+1}=\Mot(z',\mub_j)\Bigr|_{r_3(z')=0},&\qquad  j=1,\dots,b.
\end{aligned}
\ee

We see that all the columns of the obtained matrix coincide with the ones of the matrix in \eqref{FF1-ans-m1},
except the $(a+1)$-th column (associated with the parameter $z'$), where non-zero matrix elements are
 \begin{equation}\label{ext-col}
\Mot(z',\mub_j)\Bigr|_{r_3(z')=0}=t(\mub_j,z')\frac{h(\bmub,z')}{h(\bmuc,z')}
=t(z,\mub_j)\frac{g(\bmuc,z)}{g(\bmub,z)}.
 \end{equation}
It is easy to see that
 \begin{equation}\label{z-z}
\frac{-\Mot(z',\mub_j)\Bigr|_{r_3(z')=0}}{f(\bmub,z')f(z,\blab)f(\bmuc,z)}
 =\lim_{r_3(z)\to\infty}\frac1{r_3(z)}\Mot(z,\mub_j).
 \end{equation}
Thus we obtain for all the elements of the $(a+1)$-th column
 \begin{equation}\label{ext-col1}
\frac{-\tNot_{j,a+1}}{f(\bmub,z')f(z,\blab)f(\bmuc,z)}
 =\lim_{r_3(z)\to\infty}\frac1{r_3(z)}\Not_ {j,a+1},\qquad j=1,\dots,a+b+1.
 \end{equation}
Hence, we arrive at
\be{FF1-ans-m2}
\Omega_2=\lim_{r_3(z)\to\infty}\frac{\Hot_{a+1,b}}{r_3(z)}\;
\det_{a'+b}\Not .
\ee
It remains to combine \eqref{FF1-ans-m1} and \eqref{FF1-ans-m2}. This can be easily done, because
for any linear function of $\phi(\zeta)=A\zeta+B$ one has
\be{zeta}
\phi(\zeta)\Bigr|_{\zeta=0}+\zeta\lim_{\zeta\to\infty}\frac1\zeta\phi(\zeta)=\phi(\zeta).
\ee
Since the form factor $\mathcal{F}_{a,b}^{(1,2)}(z)$ is a linear function of $r_3(z)$, we obtain
\be{fin-answ}
 \mathcal{F}_{a,b}^{(1,2)}(z)=r_3(z)\Omega_2+\Omega_1=\Hot_{a+1,b}\;
\det_{a+b+1}\Not\,.
\ee

\section{Other form factors \label{S-OFF}}

Consider again the form factor of the operator $T_{12}$
 \be{SP-deFF-12pr}
\mathcal{F}_{\ta,\tb}^{(1,2)}(\tz|\blacp,\bmucp;\blabp,\bmubp)=
 \mathbb{C}^{\ta+1,\tb}(\blacp;\bmucp)T_{12}(\tz)\mathbb{B}^{\ta,\tb}(\blabp;\bmubp).
 \ee
Applying the mapping $\varphi$ \eqref{def-phi} we obtain
\be{phiFF-12}
\varphi\big(\mathcal{F}_{\ta,\tb}^{(1,2)}(\tz|\blacp,\bmucp;\blabp,\bmubp)\big)=
\mathcal{F}_{\tb,\ta}^{(2,3)}(-\tz|-\bmucp,-\blacp;-\bmubp,-\blabp)\Bigr|_{r_1\leftrightarrow r_3}.
\ee
Thus, in order to obtain the determinant representation for the
form factor $\mathcal{F}_{a,b}^{(2,3)}(z)$  one should take the resulting formulas for $\mathcal{F}_{\ta,\tb}^{(1,2)}(\tz)$,
set there
\be{replac}
\begin{array}{ll}
\ta =  b,& \quad \tb =  a,\qquad \quad  \tz =  -z\,;\\
\blacp =  -\bmuc,& \quad \bmucp =  - \blac\,;\\
\blabp =  -\bmub,& \quad  \bmubp =  - \blab\,,
\end{array}
\ee
and replace the function $r_1$ by $r_3$ and vice versa. One can say that the mapping $\varphi$ actually acts on the
determinant representation \eqref{FF-ans-m} via the replacements described above.

Consider how $\varphi$ acts on the prefactor $\Hot_{\ta+1,\tb}=\Hot_{\ta+1,\tb}(\bar{\tilde x};\blacp;\bmubp)$,
where $\bar{\tilde x}=\{\blabp,\tz,\bmucp\}$. We have
\be{Habp}
\Hot_{\ta+1,b}=\frac{h(\bar{\tilde x},\blabp)h(\bmucp,\bar{\tilde x})}{h(\bmucp,\blabp)}\;\Delta'_{\ta+1}(\blacp)
\Delta'_{b}(\bmubp)\Delta_{\ta+b+1}(\bar{\tilde x})\,.
\ee
Then
\be{PHabp1}
\varphi(\Hot_{\ta+1,\tb})=\frac{h(-\bar y,-\bmub)h(-\blac,-\bar y)}{h(-\blac,-\bmub)}\;\Delta'_{b+1}(-\bmuc)
\Delta'_{a}(-\blab)\Delta_{a+b+1}(-\bar y)\,,
\ee
where $\bar y=\{\bmub,z,\blac\}$. Since the function $g(u,v)$ possesses the property $g(-u,-v)=g(v,u)$, we
conclude that all other rational functions \eqref{univ-not}, \eqref{desand} also have similar property, and
\be{D-D}
\Delta_n(-\bw)=\Delta'_n(\bw)=(-1)^{\frac{n(n-1)}2}\Delta_n(\bw)\,.
\ee
Hence, we obtain
\be{PHabp2}
\varphi(\Hot_{\ta+1,\tb})=\frac{h(\bmub,\bar y)h(\bar y,\blac)}{h(\bmub,\blac)}\;\Delta_{b+1}(\bmuc)
\Delta_{a}(\blab)\Delta'_{a+b+1}(\bar y)=(-1)^{a(b+a)}\Htt_{a,b+1}\,.
\ee

Similarly one can convince oneself that
\be{Phi-NuNv}
\varphi\bigl(\Lot({\tilde x}_k,{\tilde u}^{\scriptscriptstyle C}_j)\bigr)= \Mtt(y_k,\muc_j)\,,\qquad
\varphi\bigl(\Mot({\tilde x}_k,{\tilde v}^{\scriptscriptstyle B}_j)\bigr)=\Ltt(y_k,\lab_j)\,.
\ee
Thus the action of the mapping $\varphi$ onto the matrix $\Not$ gives the matrix $\Ntt$ up to the
permutation of the first $b+1$ rows with the last $a$ rows. Hence,
\be{Phi-detN}
\varphi\bigl(\det\Not\bigr)=(-1)^{a(b+1)}\det\Ntt\,.
\ee
Combining \eqref{PHabp2} and \eqref{Phi-detN} we arrive at \eqref{3FF-ans-m}, and thus, we prove
Proposition~\ref{PFF-23}.

The form factors $\mathcal{F}_{a,b}^{(j+1,j)}(z)$ (with $j=1,2$) can be obtained from  $\mathcal{F}_{a,b}^{(j,j+1)}(z)$
by the mapping $\psi$ \eqref{def-psi}. On the other hand one can easily check that  making the replacements
$\{\blac,\bmuc\}\leftrightarrow\{\blab,\bmub\}$ and $\{a,b\}\leftrightarrow\{a',b'\}$ in the determinant representation
\eqref{FF-ans-m} for $\mathcal{F}_{a,b}^{(1,2)}(z)$ we arrive at \eqref{3FF-ans-m} for $\mathcal{F}_{a,b}^{(2,3)}(z)$.
Since the mapping $\psi$ yields the same replacements of the parameters, we conclude that applying $\psi$ to
$\mathcal{F}_{a,b}^{(1,2)}(z)$ and $\mathcal{F}_{a,b}^{(2,3)}(z)$ we obtain the determinant representations for
$\mathcal{F}_{a,b}^{(3,2)}(z)$ and $\mathcal{F}_{a,b}^{(2,1)}(z)$ respectively. In this way we prove
Proposition~\ref{PFF-32-21}.

\section*{Conclusion}

In this paper we considered the form factors of the monodromy matrix entries in the models with
$GL(3)$-invariant $R$-matrix. We obtained determinant representations for the form factors
$\mathcal{F}_{a,b}^{(i,j)}(z)$
of the operators $T_{ij}(z)$ with $|i-j|=1$.
In our previous publication \cite{BelPRS13a} we have already calculated the form factors of the diagonal entries
$T_{ii}(z)$. Thus, the only unknown remains the form factor $\mathcal{F}_{a,b}^{(1,3)}(z)$ of the operator $T_{13}(z)$
(the form factor $\mathcal{F}_{a,b}^{(3,1)}(z)$ can be obtained from $\mathcal{F}_{a,b}^{(1,3)}(z)$ via the mapping
$\psi$ \eqref{psiFF}). The question of whether or not there exists a determinant representation for this form factor
remains open up to now.

One of possible ways to solve this problem is quite similar to the method used in this paper. The action of the
operator $T_{13}(z)$  on the Bethe vectors is very simple
 \be{act13}
 T_{13}(z)\mathbb{B}^{a,b}(\bla;\bmu)=
 \mathbb{B}^{a+1,b+1}(\{\bla,z\};\{\bmu,z\}).
 \ee
Thus, the form factor $\mathcal{F}_{a,b}^{(1,3)}(z)$ is equal to the scalar product of the vectors
$\mathbb{C}^{a+1,b+1}(\blac;\bmuc)$ and $\mathbb{B}^{a+1,b+1}(\{\blab,z\};\{\bmub,z\})$, and we can
use the equation \eqref{Resh-SP} for the general scalar product of Bethe vectors. However, further summation
with respect to partitions of the Bethe parameters meets certain technical difficulties. In particular,
one needs to have new generalizations of the summation lemma~\ref{Ident-G}.

Another possible way to solve the problem is to use the multiple integral representation for scalar
products of the Bethe vectors obtained recently in \cite{Whe13}. This representation might be useful for
the study of the form factor $\mathcal{F}_{a,b}^{(1,3)}(z)$, if we understand how it can reproduce the
results  already obtained.

Concluding this paper we would like to say few words about possible applications of the results obtained.
Models with higher rank symmetries play an important role in condensed matter physics. They appear for instance in two-component Bose (or Fermi) gas and in the study of models of cold atoms (for e.g. ferromagnetism or phase separation). One can also mention 2-band Hubbard models (mostly in the half-filled regime), in the context of strongly correlated electronic systems. In that case, the symmetry increases when spin and orbital degrees of freedom are supposed to play a symmetrical role leading to an $SU(4)$ or even an $SO(8)$ symmetry (see e.g. \cite{su4-Hub,so8-Hub}). All these studies require to look for integrable
models with $SU(N)$ symmetry, the first step being the $SU(3)$ case. Compact determinant representations for form factors of the monodromy matrix entries give a possibility to study  correlation functions of such models. We have mentioned already in the Introduction
that these representations allow one to calculate the correlation functions of integrable spin chains via their form factor expansion.
Furthermore,  the explicit representations for the form factors also play an important role in the models, for which the solution of the inverse scattering problem is not known (see e.g. \cite{KitKMST12,CauPS07}).
In this context it is worth mentioning the work \cite{PozOK12}, where the form factors in the model of two-component Bose gas were studied.

Apart from condensed matter physics, let us also mention super-Yang-Mills theories. Integrability has proved to be a very efficient tool for the calculation of scattering amplitudes in these models. The calculation of these amplitudes  can be related to scalar products of Bethe vectors. In particular, in the $SU(3)$ subsector of the theory, one just needs the $SU(3)$ Bethe vectors. Hence, the knowledge of the form factors is also essential in this context.

Finally, in view of the potential applications, there is reason to wonder whether the results obtained in the present paper
could be generalized
to the models based on $GL(N)$ group with $N>3$. However, the structure of the obtained determinant representations  does not provide obvious clues about their possible generalization
to the case $N>3$.  We would like to be very cautious with any `obvious' predictions in this field. It is sufficient to recall some conjectures formulated previously on the basis of the results obtained in  $GL(2)$-based models. Indeed, in the case $N=2$ the analogs of the form factors considered in the present paper are proportional to the Jacobian of the transfer matrix eigenvalue on one of the vectors. The natural hypothesis was that this structure is preserved in the case $N>2$. We see, however, that already for $N=3$ the determinant representations have a more complicated structure. In particular they
contain the Jacobians of the transfer matrix eigenvalues on  both vectors.
It is very possible that in the case $N>3$, the determinant representations for form factors (if they exist) have more sophisticated structure, that is difficult to see in the case of $N=3$. Therefore we believe that the systematic study of the problem of generalization is the only way to solve it. In this context let us quote the work \cite{suN-results} where some preliminary results for $GL(N)$-based models were obtained.

\section*{Acknowledgements}
The work of S.P. was supported in part by RFBR grant 11-01-00980-a and  grant
of Scientific Foundation of NRU HSE 12-09-0064. E.R. was supported by ANR Project
DIADEMS (Programme Blanc ANR SIMI1 2010-BLAN-0120-02).
N.A.S. was  supported by the Program of RAS Basic Problems of the Nonlinear Dynamics,
RFBR-11-01-00440-a, RFBR-13-01-12405-ofi-m2, SS-4612.2012.1.

\appendix

\section{Properties of DWPF \label{A-ID}}

It follows from the representation \eqref{K-def} that the DWPF $\Izer_n(\bar x|\bar y)$ has simple poles at $x_j=y_k$. The behavior of $\Izer_n$ near  these poles can be expressed in terms of $\Izer_{n-1}$:
 \be{Rec-Ky}
\lim_{z'\to z} f^{-1}(z',z)\;\Izer_{n+1}(\{\bar x, z'\}|\{\bar y,z\})=
f(z,\bar y)f(\bar x,z)\cdot \Izer_{n}(\bar x|\bar y).
\ee

One can also easily check that the DWPF possesses the properties:
 \be{K-K}
\Izer_{n+1}(\{\bar x, z-c\}|\{\bar y, z\})=\Izer_{n+1}(\{\bar x, z\}|\{\bar y, z+c\})= - \Izer_{n}(\bar x|\bar y).
\ee
 \be{Red-K}
\Izer_{n}(\bar x-c|\bar y)=\Izer_{n}(\bar x|\bar y+c)= (-1)^n f^{-1}(\bar y,\bar x) \Izer_{n}(\bar y|\bar x).
\ee

The DWPF $\Izer_n$ satisfies several summation identities.
\begin{lemma}\label{main-ident}
Let $\bar\xi$, $\bar\alpha$ and $\bar\beta$ be sets of complex variables with $\#\alpha=m_1$,
$\#\beta=m_2$, and $\#\xi=m_1+m_2$. Then
\begin{equation}\label{Sym-Part-old1}
  \sum
 \Izer_{m_1}(\bar\xi_{\so}|\bar \alpha)\Izer_{m_2}(\bar \beta|\bar\xi_{\st})f(\bar\xi_{\st},\bar\xi_{\so})
 = (-1)^{m_1}f(\bar\xi,\bar \alpha) \Izer_{m_1+m_2}(\{\bar \alpha-c,\bar \beta\}|\bar\xi).
 \end{equation}
The sum is taken with respect to all partitions of the set $\bar\xi$ into
subsets $\bar\xi_{\so}$ and $\bar\xi_{\st}$ with $\#\bar\xi_{\so}=m_1$ and $\#\bar\xi_{\st}=m_2$.
\end{lemma}

\begin{lemma}\label{Long-Det}
Let $\bar \gamma$ and $\bar\xi$ be two sets of generic complex numbers with $\#\bg=\#\bar\xi=m$. Let
also $C_1(\gamma)$ and $C_2(\gamma)$ be two arbitrary functions of a complex variable $\gamma$. Then
\begin{multline}\label{SumDet1}
\sum \Izer_m(\{\bg_{\so}-c, \bg_{\st}\}|\bar \xi)f(\bar \xi, \bg_{\so})f(\bg_{\st},\bg_{\so})
C_1(\bg_{\so})C_2(\bg_{\st})\num
=\Delta'_m(\bar\xi)\Delta_m(\bg)
\det_m\Bigl(C_2(\gamma_k)t(\gamma_k,\xi_j)h(\gamma_k,\bar\xi)+(-1)^m C_1(\gamma_k)t(\xi_j,\gamma_k)h(\bar\xi,\gamma_k)\Bigr).
\end{multline}
Here we use the shorthand  notation \eqref{SH-prod} for the products of the functions $C_1$ and $C_2$.
\end{lemma}

The proofs of these lemmas is given in \cite{BelPRS12b}.

\begin{lemma}\label{Wau}
Let $\bar\alpha$ and $\bar\beta$ be two sets of generic complex numbers with $\#\bar\alpha=\#\bar\beta=m$,
and $z$ is an arbitrary complex.
Then
\begin{multline}\label{Ident-G}
\sum_{\substack{\bar\alpha=\{\bar\alpha_{\so},\;\bar\alpha_{\st}\} \\ \bar\beta=\{\bar\beta_{\so},\;\bar\beta_{\st}\}}}
  f(\bar\beta_{\so},z)f(\bar\beta_{\st},\bar\beta_{\so})f(\bar\alpha_{\so},\bar\alpha_{\st})   \Izer_{m_{\so}}(\bar\beta_{\so}|\bar\alpha_{\so})\Izer_{m_{\st}}(\bar\alpha_{\st}+c|\bar\beta_{\st})\num
  =(-1)^m t(\bar\alpha,\bar\beta)h(\bar\alpha,\bar\alpha)h(\bar\beta,\bar\beta)h(\bal,z)g(\bbet,z),
       \end{multline}
where the sum is taken over all possible partitions of the sets $\bar\alpha$ and $\bar\beta$
with $\#\bar\alpha_{\so}=\#\bar\beta_{\so}=m_{\so}$, $m_{\so}=0,\dots,m$, and $\#\bar\alpha_{\st}=\#\bar\beta_{\st}=m_{\st}=m-m_{\so}$.
\end{lemma}
This lemma is a generalization of the lemma~6.3 of the work \cite{BelPRS12b}. In particular, the statement of the
latter can be obtained from \eqref{Ident-G} in the limit $z\to\infty$.

{\sl Proof.}  Let us denote by $\Lambda^{(l)}_m(\bar\alpha|\bar\beta)$ and $\Lambda^{(r)}_m(\bar\alpha|\bar\beta)$
the l.h.s. and the r.h.s. of \eqref{Ident-G} respectively. Obviously, they are symmetric functions of $\bar\alpha$ and symmetric functions of $\bar\beta$. Consider them as functions of $\alpha_1$, while
all other variables are fixed.  Clearly the functions $\Lambda^{(l)}_m$ and $\Lambda^{(r)}_m$ are rational functions of $\alpha_1$,
and they both vanish if  $\alpha_1\to\infty$. They have poles at $\alpha_1=\beta_k$ and $\alpha_1+c=\beta_k$, $k=1,\dots,m$. Some
terms in the l.h.s. of \eqref{Ident-G} may also have  poles at $\alpha_1=\alpha_k$, $k=2,\dots,m$, but it is not difficult to
show that these singularities cancel each other.  Indeed, the sum over partitions $\bar\alpha\Rightarrow\{\bar\alpha_{\so},\bar\alpha_{\st}\}$ can be explicitly calculated via \eqref{Sym-Part-old1}
\be{sum-alpha}
\sum_{\bar\alpha=\{\bar\alpha_{\so},\;\bar\alpha_{\st}\} }
f(\bar\alpha_{\so},\bar\alpha_{\st})   \Izer_{m_{\so}}(\bar\beta_{\so}|\bar\alpha_{\so})\Izer_{m_{\st}}(\bar\alpha_{\st}+c|\bar\beta_{\st})=
(-1)^{m_{\st}}f(\bar\alpha,\bar\beta_{\st}-c)\Izer_m(\{\bar\beta_{\st}-2c,\bar\beta_{\so}\}|\bar\alpha),
\ee
and we see that the r.h.s. of \eqref{sum-alpha} is well defined at  $\alpha_1=\alpha_k$. Finally, it is easy to check that
 \be{G1}
 \Lambda^{(l)}_1(\alpha_1|\beta_1)=\Lambda^{(r)}_1(\alpha_1|\beta_1)=-t(\alpha_1,\beta_1)h(\alpha_1,z)g(\beta_1,z).
 \ee

The listed properties allow us to prove \eqref{Ident-G} via induction over $m$.  We assume that it holds
for $\#\bar\alpha=\#\bar\beta=m-1$  and consider the case  $\#\bar\alpha=\#\bar\beta=m$. Let us calculate
the residues of $\Lambda^{(l)}_m(\bar\alpha|\bar\beta)$ and $\Lambda^{(r)}_m(\bar\alpha|\bar\beta)$ at
$\alpha_1=\beta_k$ and $\alpha_1+c=\beta_k$. Obviously, due to the symmetry of
$\Lambda^{(l)}_m$ and $\Lambda^{(r)}_m$ over $\bar\beta$ it is enough to consider $\beta_k=\beta_1$.

It is straightforward to establish the following recursions:
\begin{align}\label{recursL1}
 \Bigl. \Lambda^{(r)}_m(\bar\alpha|\bar\beta)\Bigr|_{\alpha_1\to \beta_1} &=
 -g(\alpha_1, \beta_1)f(\alpha_1,z)f(\bar\beta_1,\beta_1)f(\alpha_1,\bar\alpha_1)\cdot \Lambda^{(r)}_{m-1}(\bar\alpha_1|\bar\beta_1),\\
 \Bigl. \Lambda^{(r)}_m(\bar\alpha|\bar\beta)\Bigr|_{\alpha_1+c \to \beta_1} &=
 h^{-1}(\alpha_1, \beta_1)f(\beta_1,\bar\beta_1)f(\bar\alpha_1,\alpha_1)\cdot \Lambda^{(r)}_{m-1}(\bar\alpha_1|\bar\beta_1),
 \label{recursL2}
\end{align}
where $\bar\alpha_1=\bar\alpha\setminus\alpha_1$
and $\bar\beta_1=\bar\beta\setminus\beta_1$. Let us check that $\Lambda^{(l)}_m(\bar\alpha|\bar\beta)$
has the same recursion properties.

Consider, for example, the pole at $\alpha_1=\beta_1$. This pole appears if and only if
$\alpha_1\in\bar\alpha_{\so}$ and $\beta_1\in\bar\beta_{\so}$. Let $\bar\alpha_{{\so}'}=\bar\alpha_{\so}\setminus\alpha_1$
and $\bar\beta_{{\so}'}=\bar\beta_{\so}\setminus\beta_1$.
Then using the recursion properties of the DWPF we obtain
\begin{multline}\label{G2-pole}
\Bigl. \Lambda^{(l)}_m(\bar\alpha|\bar\beta)\Bigr|_{\alpha_1\to \beta_1}
={\sum}'  f(\alpha_{1},z) f(\bar\beta_{{\so}'},z) f(\bar\beta_{\st},\bar\beta_{{\so}'})f(\bar\beta_{\st},\beta_{1})
 f(\bar\alpha_{{\so}'},\bar\alpha_{\st}) f(\alpha_{1},\bar\alpha_{\st})
  \\
 \times
 g(\beta_1,\alpha_1)f(\alpha_{1},\bar\alpha_{{\so}'})f(\bar\beta_{{\so}'},\beta_{1})
 \Izer_{m_{\so}-1}(\bar\beta_{{\so}'}|\bar\alpha_{{\so}'})\Izer_{m_{\st}}(\bar\alpha_{\st}+c|\bar\beta_{\st}),
       \end{multline}
where ${\sum}'$ means that the sum is taken over partitions of the sets $\bar\alpha_1$ and $\bar\beta_1$. Obviously
\be{comb}
\begin{array}{l}
f(\alpha_{1},\bar\alpha_{\st})f(\alpha_{1},\bar\alpha_{{\so}'})=f(\alpha_{1},\bar\alpha_1),\\
f(\bar\beta_{\st},\beta_{1})f(\bar\beta_{{\so}'},\beta_{1})=f(\bar\beta_1,\beta_{1}),
\end{array}
\ee
and thus, these factors can be moved out off the sum over partitions. We obtain
%
\begin{eqnarray}\label{G2-pole2}
\Bigl. \Lambda^{(l)}_m(\bar\alpha|\bar\beta)\Bigr|_{\alpha_1\to \beta_1}
&=&g(\beta_1,\alpha_1)f(\alpha_{1},\bar\alpha_1)f(\bar\beta_1,\beta_{1})f(\alpha_{1},z)
\nonumber \\
&& \times {\sum}'  f(\bar\beta_{{\so}'},z)f(\bar\beta_{\st},\bar\beta_{{\so}'})
 f(\bar\alpha_{{\so}'},\bar\alpha_{\st})
  \Izer_{m_{\so}-1}(\bar\beta_{{\so}'}|\bar\alpha_{{\so}'})\Izer_{m_{\st}}(\bar\alpha_{\st}+c|\bar\beta_{\st})
  \nonumber\\
 & =&-g(\alpha_1,\beta_1)f(\alpha_{1},\bar\alpha_1)f(\bar\beta_1,\beta_{1})f(\alpha_{1},z)
  \cdot \Lambda^{(l)}_{m-1}(\bar\alpha_1|\bar\beta_1).
\end{eqnarray}

Let now $\alpha_1+c=\beta_1$. Then the pole appears if and only if
$\alpha_1\in\bar\alpha_{\st}$ and $\beta_1\in\bar\beta_{\st}$. Let $\bar\alpha_{{\st}'}=\bar\alpha_{\st}\setminus\alpha_1$
and $\bar\beta_{{\st}'}=\bar\beta_{\st}\setminus\beta_1$.
Then  we obtain
\begin{eqnarray}\label{G2-pole+}
\Bigl. \Lambda^{(l)}_m(\bar\alpha|\bar\beta)\Bigr|_{\alpha_1+c\to \beta_1}
&=&{\sum}'    f(\bar\beta_{\so},z) f(\bar\beta_{1},\bar\beta_{\so}) f(\bar\beta_{\st'},\bar\beta_{\so})
 f(\bar\alpha_{\so},\bar\alpha_{\st'}) f(\bar\alpha_{\so},\alpha_{1})
\nonumber\\
&&\qquad \times
 g(\alpha_1+c,\beta_1)f(\bar\alpha_{{\st}'},\alpha_{1})f(\beta_{1},\bar\beta_{{\st}'})
 \Izer_{m_{\so}}(\bar\beta_{{\so}}|\bar\alpha_{{\so}})\Izer_{m_{\st}-1}(\bar\alpha_{\st'}+c|\bar\beta_{\st'})
\nonumber \\
 &=&h^{-1}(\alpha_1, \beta_1)f(\beta_1,\bar\beta_1)f(\bar\alpha_1,\alpha_1)
 \cdot \Lambda^{(l)}_{m-1}(\bar\alpha_1|\bar\beta_1).
\end{eqnarray}
 Thus, due to the induction assumption the residues of the functions $\Lambda^{(l)}_m(\bar\alpha|\bar\beta)$ and $\Lambda^{(r)}_m(\bar\alpha|\bar\beta)$ in the poles at $\alpha_1=\beta_1$ and $\alpha_1+c=\beta_1$ coincide. Hence,
the difference $\Lambda^{(l)}_m(\bar\alpha|\bar\beta)-\Lambda^{(r)}_m(\bar\alpha|\bar\beta)$ is a holomorphic
function of $\alpha_1$ in the whole complex plane. Since this function vanishes at $\alpha_1\to\infty$, we conclude that
$\Lambda^{(l)}_m(\bar\alpha|\bar\beta)=\Lambda^{(r)}_m(\bar\alpha|\bar\beta)$.\qed

\begin{cor}\label{Wau1}
At the conditions of lemma~\ref{Wau}
\begin{multline}\label{Ident-G1}
\sum_{\substack{\bar\alpha=\{\bar\alpha_{\so},\;\bar\alpha_{\st}\} \\ \bar\beta=\{\bar\beta_{\so},\;\bar\beta_{\st}\}}}
  f^{-1}(\bar\beta_{\st},z)f(\bar\beta_{\st},\bar\beta_{\so})f(\bar\alpha_{\so},\bar\alpha_{\st})   \Izer_{m_{\so}}(\bar\beta_{\so}|\bar\alpha_{\so})\Izer_{m_{\st}}(\bar\alpha_{\st}+c|\bar\beta_{\st})\num
  =(-1)^m t(\bar\alpha,\bar\beta)h(\bar\alpha,\bar\alpha)h(\bar\beta,\bar\beta)\frac{h(\bal,z)}{h(\bbet,z)}.
       \end{multline}
\end{cor}
{\sl Proof.} Dividing both sides of \eqref{Ident-G} by $f(\bar\beta,z)$ we immediately
arrive at \eqref{Ident-G1}.

\begin{cor}\label{Wau2}
Let $\bar\alpha$, $\bar\beta$, and $z$ be  generic complex numbers with $\#\bar\alpha=m$ and
$\#\bar\beta=m-1$. Then
\begin{multline}\label{Ident-G2}
\sum_{\substack{\bar\alpha=\{\bar\alpha_{\so},\;\bar\alpha_{\st}\} \\ \bar\beta=\{\bar\beta_{\so},\;\bar\beta_{\st}\}}}
  f(\bar\beta_{\st},\bar\beta_{\so})f(\bar\alpha_{\so},\bar\alpha_{\st})   \Izer_{m_{\so}}(\{z,\bar\beta_{\so}\}|\bar\alpha_{\so})\Izer_{m_{\st}}(\bar\alpha_{\st}+c|\bar\beta_{\st})\num
  =(-1)^m t(\bar\alpha,\bar\beta)h(\bar\alpha,\bar\alpha)h(\bar\beta,\bar\beta)g(\bal,z)h(z,\bbet).
       \end{multline}
where the sum is taken over all possible partitions of the sets $\bar\alpha$ and $\bar\beta$
with $\#\bar\alpha_{\so}=\#\bar\beta_{\so}+1=m_{\so}$, $m_{\so}=1,\dots,m$, and $\#\bar\alpha_{\st}=\#\bar\beta_{\st}=m_{\st}=m-m_{\so}$.
\end{cor}

{\sl Proof.} Consider the residue of \eqref{Ident-G} at $\beta_1=z$. We have in the r.h.s.
\be{Rbet1-z}
\Bigl. \Lambda^{(r)}_m(\bar\alpha|\bar\beta)\Bigr|_{\beta_1\to z}=g(\beta_1, z)f(\bbet_1,z)\cdot
(-1)^m t(\bar\alpha,\bar\beta_1)h(\bar\alpha,\bar\alpha)h(\bar\beta_1,\bar\beta_1)g(\bal,z)h(z,\bbet_1).
\ee

In the l.h.s. the pole occurs if and only if $\beta_1\in\bbet_{\so}$. Setting $\bar\beta_{{\so}'}=
\bbet_{\so}\setminus\beta_1$, we obtain

\begin{multline}\label{Lbet1-z}
\Bigl. \Lambda^{(l)}_m(\bar\alpha|\bar\beta)\Bigr|_{ \beta_1\to z}
={\sum}'  g(\beta_1,z)f(\bar\beta_{{\so}'},z) f(\bar\beta_{\st},\bar\beta_{{\so}'})f(\bar\beta_{\st},z)
 f(\bar\alpha_{\so},\bar\alpha_{\st})
  \\
 \times
 \Izer_{m_{\so}}(\{z,\bar\beta_{{\so}'}\}|\bar\alpha_{\so})\Izer_{m_{\st}}(\bar\alpha_{\st}+c|\bar\beta_{\st}),
 \end{multline}
where ${\sum}'$ means that the sum is taken over partitions of the sets $\bar\alpha$ and $\bar\beta_1$.
Using
\be{comb-3}
f(\bar\beta_{{\so}'},z)f(\bar\beta_{\st},z)=f(\bar\beta_1,z),
\ee
we obtain
 \begin{equation}\label{Lbet1-z2}
\Bigl. \Lambda^{(l)}_m(\bar\alpha|\bar\beta)\Bigr|_{ \beta_1\to z} =g(\beta_1,z)f(\bar\beta_1,z){\sum}'   f(\bar\beta_{\st},\bar\beta_{{\so}'})
 f(\bar\alpha_{\so},\bar\alpha_{\st})\;
  \Izer_{m_{\so}}(\{z,\bar\beta_{{\so}'}\}|\bar\alpha_{\so})\Izer_{m_{\st}}(\bar\alpha_{\st}+c|\bar\beta_{\st}).
       \end{equation}
Comparing \eqref{Lbet1-z2} and \eqref{Rbet1-z} we arrive at the statement of Corollary~\ref{Wau2}.

\end{document}